\documentclass[fleqn,usenatbib]{mnras}

\usepackage{newtxtext,newtxmath}

\usepackage[T1]{fontenc}

\DeclareRobustCommand{\VAN}[3]{#2}
\let\VANthebibliography\thebibliography
\def\thebibliography{\DeclareRobustCommand{\VAN}[3]{##3}\VANthebibliography}

\usepackage{graphicx}	
\usepackage{subcaption}
\usepackage{amsmath}	
\usepackage{caption}  
\usepackage{subcaption}

\newcommand{\Rey}[1]{\overline{#1}}
\newcommand{\Fav}[1]{\widetilde{#1}}

\title[Complete Convective Silicon Shell in 3D]{3D simulations of a complete convective silicon shell burning phase}

\author[V. Varma et al. 2026]{V. Varma$^{1}$\thanks{E-mail: v.r.vejayan@keele.ac.uk}, R. Hirschi$^{1,2}$, F. Rizzuti$^{3,4,5}$, T. Rauscher$^{6,7}$, A. St.J. Murphy$^{8}$, M. Moc\'ak$^{9}$, C. Meakin$^{9}$, 
\newauthor K. Goodman$^{1}$, C. Georgy$^{10}$ and W.D. Arnett$^{11}$
\\
$^{1}$Astrophysics Research Centre, Lennard-Jones Laboratories, Keele University, Keele ST5 5BG, UK\\
$^{2}$Kavli IPMU (WPI), University of Tokyo, 5-1-5 Kashiwanoha, Kashiwa 277-8583, Japan\\
$^{3}$Heidelberger Institut f{\"u}r Theoretische Studien, Schloss-Wolfsbrunnenweg 35, D-69118 Heidelberg, Germany\\
$^{4}$INAF, Osservatorio Astronomico di Trieste, via G.B. Tiepolo 11, I-34131 Trieste, Italy\\
$^{5}$INFN, Sezione di Trieste, via Valerio 2, I-34134 Trieste, Italy\\
$^{6}$Department of Physics, University of Basel, 4056 Basel, Switzerland\\
$^{7}$Centre for Astrophysics Research, University of Hertfordshire, Hatfield AL10 9AB, UK\\
$^{8}$School of Physics and Astronomy, University of Edinburgh, Edinburgh EH9 3FD, UK\\
$^{9}$Aster Scientific, Pasadena, CA 91104 USA \\
$^{10}$Geneva Observatory, Geneva University, CH-1290 Sauverny, Switzerland \\
$^{11}$Steward Observatory, University of Arizona, 933 N. Cherry Avenue, Tucson AZ 85721, USA}

\date{Accepted XXX. Received YYY; in original form ZZZ}

\pubyear{2024}

\begin{document}
\label{firstpage}
\pagerange{\pageref{firstpage}--\pageref{lastpage}}
\maketitle

\begin{abstract}
We present 3D hydrodynamic simulations of a complete silicon shell burning phase until fuel exhaustion at the end of the evolution of a 14$M_\odot$ core-collapse supernova progenitor, using a reduced 25-isotope nuclear network. We investigate how realistic turbulent convection affects this burning phase, which has a more complicated set of nuclear reactions than previous burning phases. We find broad similarities between the 3D simulation and the 1D \textsc{MESA} model. However, due to more realistic feedback from the very stiff convective boundaries in the hydrodynamic simulations, the 3D simulation experiences lower convective boundary mixing (CMB) compared to 1D, and hence entrains less fresh fuel into the silicon shell. This leads to the silicon shell in the 3D model burning for roughly 800\,s shorter. We find that the nuclear burning timescales for the dominant reactions are faster than the mixing timescale, making this entire process a convective-reactive event. The angular-averaged energy generation profile shows a double-peaked structure, where the region between the positive peaks are close to zero, or are negative. We find that throughout the base of this silicon shell, many $(\alpha, p)$ and their inverse reactions are important. The forward and reverse rates are very similar, so slight fluctuations in the temperature cause regions to be either exoergic or endoergic, leading to a complicated energy generation evolution. This study presents an exploration using a single reduced nuclear network, however, due to the sensitivity of this burning phase future studies should investigate the impact of more complete nuclear networks.
\end{abstract}




\section{Introduction}

Massive stars end their lives as in an iron core-collapse after a sequence of nuclear burning stages has transformed their interiors into highly stratified structures. By the onset of collapse, the star consists of an iron-group core surrounded by compositionally distinct shells produced by previous core and shell burning phases of silicon, oxygen, neon, carbon, helium, and hydrogen (however, see \citet{Whitehead2026} concerning shell interactions and its impact on the final onion-like structure). The details of these burning events determine the mass distribution, entropy profile, electron fraction, and composition encountered by the collapsing core and by the supernova shock. As a result, the final pre-supernova structure plays a central role in setting both the conditions for explosion and the nucleosynthetic outcome \citep{Muller2020LRA, Boccioli2024}.

Silicon burning plays a particularly important role because it is the final hydrostatic burning phase before core-collapse. It governs the late growth of the iron core and helps establish the final state of the surrounding silicon-rich material. The resulting iron-core mass, electron-fraction profile, entropy structure, and density gradient outside the core influence the subsequent accretion history during collapse, the ram pressure opposing shock expansion, and the mass coordinates through which explosive burning proceeds. Understanding silicon burning is therefore essential for connecting the late evolution of massive stars to both core-collapse supernova dynamics and its nucleosynthetic yields that are deposited to the interstellar medium \citep{Woosley2002, Nomoto2013}.

Unlike many of the early hydrostatic stellar burning phases, silicon burning is challenging to model because it is not well described as a simple linear sequence of fusion reactions. At the temperatures characteristic of hydrostatic silicon burning, typically 3-4\,GK, photodisintegration, capture reactions, and weak interactions operate simultaneously, driving the composition toward a state of partial nuclear equilibrium. Rather than evolving independently, the nuclei become linked, often into two quasi-statistical equilibrium (QSE) groups, one focused around silicon and the other around the iron peak nuclei \citep{Bodansky1968, Thielemann1985,Hix1996, Hix1998, Hix1999}. Within these groups, rapid strong and electromagnetic reactions maintain approximate equilibrium while slower reactions control the exchange between groups and the gradual approach toward nuclear statistical equilibrium (NSE). The final composition and energy generation are therefore sensitive to the thermodynamic trajectory, the electron fraction, and the degree of neutronization. In particular, changes in neutron excess can strongly modify the QSE structure and the production of neutron-rich iron-group isotopes. Accurately following this nucleosynthesis requires large nuclear networks, but their cost is prohibitive in multidimensional hydrodynamic simulations, motivating the use of reduced well-considered networks that reproduce the dominant energetics and abundance evolution while retaining the essential QSE physics \citep{Hix2007} and most recently using neural networks to reduce computational costs \citep{Grichener2025}.

Recent multidimensional simulations have demonstrated that late-stage convective shell burning can produce large-amplitude and high Mach number, non-spherical perturbations, which can aid the neutrino driven CCSNe explosion mechanism \citep{couch_14,Mueller2015, Mueller_2017}. This has led to an increase in the use of 3D simulations covering the final evolution to core-collapse. These studies have largely focused on the final convective oxygen burning shells, as these shells tend to be larger and have higher velocity flows \citep{Collins2018}. Several 3D studies have also followed the evolution of the silicon shell \citep{Yoshida2021, Fields2021}, however, their primary emphasis has been on the hydrodynamic properties of the convective shells: turbulent velocities, convective Mach numbers, dominant angular scales, shell morphology, and the consequences of these perturbations for shock revival and explosion dynamics. The study by \citet{Zingale2024} on the other hand, focuses on the transition to NSE. Although there have been some published studies that have followed the hydrostatic burning of the silicon burning shell \citep{Bazan1997, Arnett2011, chatzopoulos_16}, the impact of fully three-dimensional turbulent mixing on the evolution of the silicon burning abundance has remained comparatively unexplored. 

Convective-reactive burning occurs when the nuclear burning timescale is comparable to or faster than turbulent mixing. This regime had been studied initially in low-mass star hydrogen-ingestion events \citep{Herwig2011, Stephens2021}, where 3D simulations show that asymmetric convective transport can alter i-process nucleosynthesis and cannot be fully captured by one-dimensional mixing prescriptions. Similar behaviour appears in massive-stars, especially in ingestion events or entire shell mergers \citep{Ritter2018,Andrassy2020,Rizzuti2024}, where a sudden influx of light elements as fuel can contribute significantly to the luminosity and affect the subsequent nucleosynthesis. Silicon shell burning is therefore part of a wider class of convective-reactive events, although its relevant reactions are distinct, and it does not require an interaction with other shells.

In this study, we present a set of 3D hydrodynamic simulations of convective silicon shell burning until the exhaustion of fuel. Our paper is structured as follows: In Section~\ref{sec:setup}, we describe the initial progenitor model as well as numerical methods of the \textsc{PROMPI} code used in our simulations. This is followed by the results of these simulations. We first analyse the differences between the initial 1D \textsc{MESA} evolution compared to the 3D simulation in Section~\ref{subsec:1D_3D}, followed by comparing resolution and opening angle effects of our spherical wedge models in Section~\ref{subsec:resolution}. In Section~\ref{subsec:velocity_field} we analyse the velocity field and convective boundary mixing in the 3D simulations, before finally discussing the insights from the silicon burning process itself and the subsequent energy generation in Section~\ref{subsec:enuc}. We summarise our results and discuss their implications in Section~\ref{sec:conclusion}.

\section{Methods and Simulation Setup}
\label{sec:setup}

\begin{figure*}
\centering
    \subfloat{\includegraphics[width=0.5\linewidth]{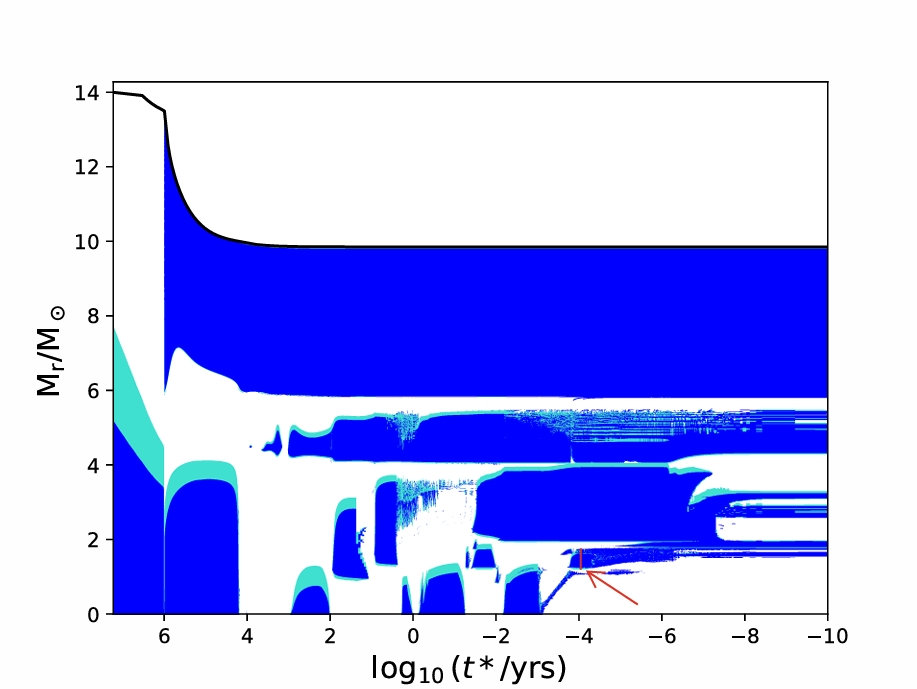}}
\hfil
    \subfloat{\includegraphics[width=0.5\linewidth]{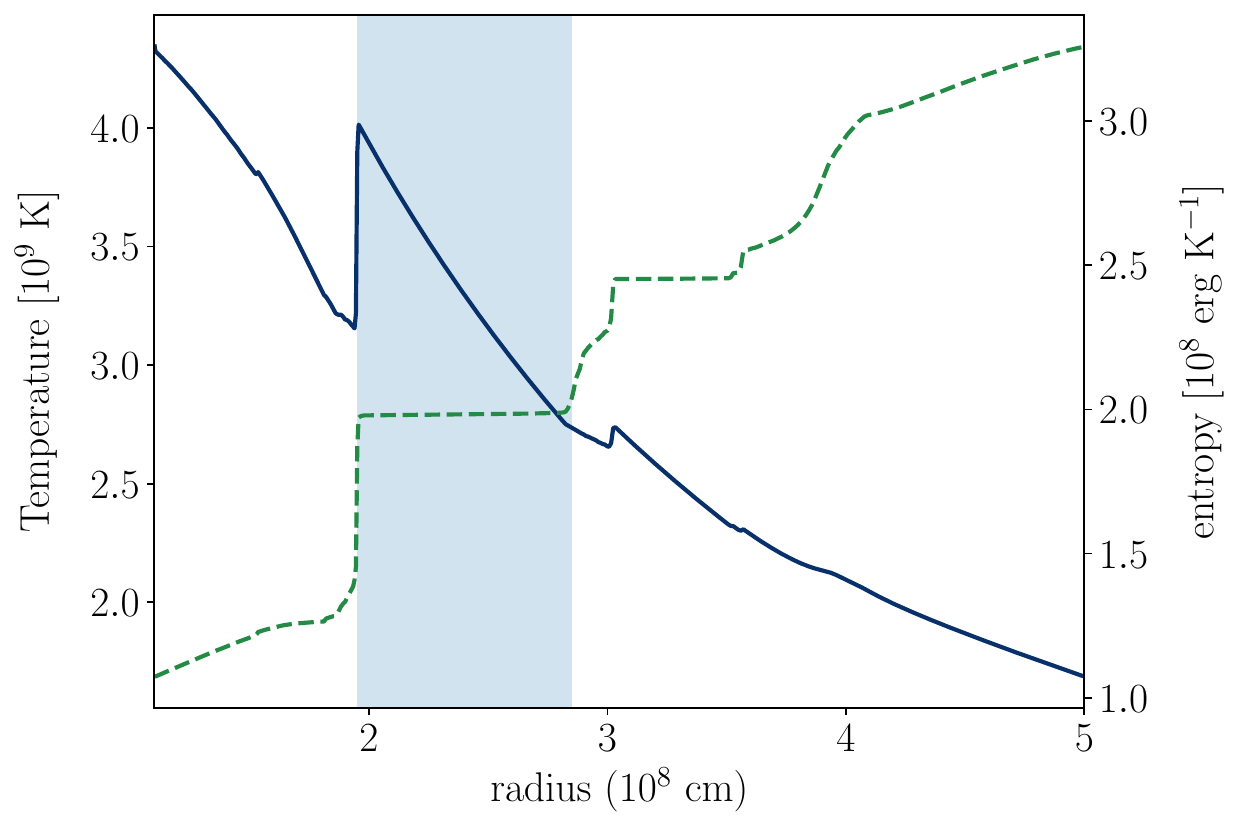}}
\caption{\normalsize Structure evolution (Kippenhahn) diagram showing the full evolution of the $14\mathrm{M_\odot}$ star (left), along with the initial temperature (solid blue) and entropy (dashed green) profiles of the region mapped to 3D (right). The primary convective silicon shell is depicted by the blue shaded regions, with the CBM shown in cyan. The red arrow indicates the silicon shell we simulate, and the vertical red line indicates where the 3D simulation starts. }
    \label{fig:init}
\end{figure*}

We simulate convective silicon-shell burning in a non-rotating 
$14\,\mathrm{M_\odot}$ massive star with an initial metallicity of Z=0.001. This model was run with  the
one-dimensional stellar evolution code \textsc{MESA} \citep{Paxton2011,Paxton2019}. 
This progenitor model was published in \citet{Whitehead2026} and is referred to as 14SH21 in that paper as it uses the mass dependent convective boundary mixing (CBM) described in \citet{Scott2021} (SH21). The full structure evolution of 
the model is shown in the left panel of 
Figure~\ref{fig:init}. We map the 1D profile to our 3D hydrodynamics code around 4000\,s before core-collapse, which corresponds to $\mathrm{log_{10}(t/yrs})=-4$ in this figure.

At this time, the main convective silicon shell extends 
from $r\simeq1.9\times10^8\,\mathrm{cm}$ to 
$r\simeq2.7\times10^8\,\mathrm{cm}$, and is bounded above and below by 
stably stratified layers. We map the full silicon-burning 
convection zone, together with a substantial fraction of the neighbouring 
radiative regions, allowing us to follow both the internal evolution of the burning shell and any convective boundary mixing that develops during the three-dimensional simulations in addition to leaving space for any possible expansion/contraction of the shell.

The right panel of Figure~\ref{fig:init} shows the initial temperature and entropy 
profiles mapped to 3D. The silicon-burning convection zone corresponds 
to a relatively flat entropy region, which we have marked with a shaded band in the figure. The initial model also contains a second, nearly flat entropy region 
above the active silicon shell, associated with an earlier convective episode 
in the one-dimensional evolution. As discussed later in this paper, the 
subsequent treatment of this region differs between the one-dimensional and 
three-dimensional calculations and plays an important role in the comparison 
between them.

We perform the three-dimensional simulations with the MPI-parallelised stellar 
hydrodynamics code \textsc{PROMPI} \citep{Meakin_2007}. \textsc{PROMPI} is 
derived from the \textsc{PROMETHEUS} code and solves the inviscid Euler 
equations using a finite-volume, Eulerian implementation of the piecewise 
parabolic method \citep[PPM;][]{Colella1984, Fryxell2000}. The code comparison study of \cite{Andrassy2022} showed that \texttt{PROMPI} is fully consistent with other hydrodynamic codes commonly employed for stellar studies.
The code has previously
been used extensively to study turbulent convection and convective boundary 
mixing in the late stages of massive-star evolution, including simulations of 
oxygen-, carbon-, and neon-burning shells \citep[e.g.][]{Meakin_2007,arnett_09,cristini_17,cristini_19,rizzuti_23,Georgy2024,Rizzuti2024}. The present calculations extend this approach to an active silicon-burning shell.

We list the complete set of combustive Euler equations solved by \texttt{PROMPI} here. A more detailed description can be found in \cite{Meakin_2007}. We present the equations in state-vector form with \textbf{\textit{Q}} the state vector, $\mathbf{\Phi}$ the flux vector, and \textbf{\textit{S}} the source vector:
\begin{equation}
\begin{aligned}
&\frac{\partial{\textbf{\textit{Q}}}}{\partial{t}}+\nabla\cdot\mathbf{\Phi} = \textbf{\textit{S}}\\
&\textbf{\textit{Q}} = \begin{cases}
        \rho \\
        \rho \textbf{\textit{v}}\\
        \rho E\\
        \rho X_i
    \end{cases}
\mathbf{\Phi} = \begin{cases}
\rho \textbf{\textit{v}}\\
\rho \textbf{\textit{v}}\cdot\textbf{\textit{v}} + p\\
(\rho E + p)\textbf{\textit{v}}\\
\rho X_i \textbf{\textit{v}}
\end{cases}
\textbf{\textit{S}} = \begin{cases}
        0\\
\rho \textbf{\textit{g}}\\
\rho \textbf{\textit{v}} \cdot \textbf{\textit{g}} + \rho \epsilon\\
R_i\\
    \end{cases}
\end{aligned}
\end{equation}
where $\rho, p, \textbf{\textit{v}}, \textbf{\textit{g}},$ and $T$ are the density, pressure, velocity, gravity, and temperature, respectively. We note that  \textbf{\textit{g}} is a time-dependent quantity that is re-evaluated at each timestep. $E$ is the total specific energy, and the energy source term $\epsilon$ is due to nuclear reactions and neutrino cooling. $R_i$ is the time rate of change of species $X_i$ due to nuclear reactions. The set of equations are closed using the tabulated `Timmes' equation of state \citep{Timmes_1999}.

The simulations are performed in spherical polar coordinates 
$(r,\theta,\phi)$. The hydrodynamic domain extends radially from 
$r_{\rm in}=1.1\times10^8\,\mathrm{cm}$ to 
$r_{\rm out}=5.0\times10^8\,\mathrm{cm}$, substantially larger than the initial 
radial extent of the convective shell itself. The full radial extent of the simulated domain is shown in the right plot of Figure~\ref{fig:init}.
This choice ensures that the 
active burning region remains well separated from the radial boundaries and 
allows for possible growth of the convective shell beyond the extent 
predicted by the one-dimensional model. The inner and outer radial boundaries 
are placed in stably stratified regions. A thin damping layer is applied near 
the outer boundary in order to reduce the reflection of gravity waves back into 
the computational domain. This prevents artificial wave reflection from 
influencing the long-term evolution of the silicon-burning shell.

Nuclear burning is followed using an approximate 25-isotope network, based on 
the network also used in \citet{Meakin2006}. The evolved species are:
n, p, $\mathrm{^4He}$, $\mathrm{^{12}C}$, $\mathrm{^{16}O}$, $\mathrm{^{20}Ne}$, $\mathrm{^{23}Na}$, $\mathrm{^{24}Mg}$, $\mathrm{^{28}Si}$, $\mathrm{^{31}P}$, $\mathrm{^{32}S}$, $\mathrm{^{34}S}$, $\mathrm{^{35}Cl}$, $\mathrm{^{36}Ar}$, $\mathrm{^{38}Ar}$, $\mathrm{^{39}K}$, $\mathrm{^{40}Ca}$, $\mathrm{^{42}Ca}$, $\mathrm{^{44}Ti}$, $\mathrm{^{46}Ti}$, $\mathrm{^{48}Cr}$, $\mathrm{^{50}Cr}$, $\mathrm{^{52}Fe}$, $\mathrm{^{54}Fe}$, $\mathrm{^{56}Ni}$. 
This network is designed to capture the dominant energetics and broad abundance 
evolution of advanced burning while remaining computationally feasible in 
three-dimensional hydrodynamic simulations. It does not provide the same level 
of nucleosynthetic detail as a large post-processing network. In particular, 
the final iron-group composition and electron-fraction evolution should be 
interpreted with the limitations of the reduced network in mind. We will comment on the processes throughout this paper where the reduced network size is likely to have an impact. 

We run simulations over two angular domain sizes: 
$45^\circ\times45^\circ$ and $90^\circ\times90^\circ$ in 
$(\theta,\phi)$. The smaller wedge provides higher effective angular 
resolution at fixed computational cost, while the larger wedge allows larger 
angular-scale convective modes to develop. 

The set of simulations presented in this paper is summarised in 
Table~\ref{tab:Models}. The models differ in angular extent, grid resolution, 
and total simulated physical time. The naming convention indicates the angular extent in degrees and resolution. Low, med and high in the run names refer to 3 different local resolution choices.

\begin{table}
    \centering
    \begin{tabular}{lccc}
    \hline
    \textbf{Model} & \textbf{Resolution} 
    & \textbf{Angular domain} $(\theta\times\phi)$ 
    & $\mathbf{t_{\rm sim}}$ \\
    \hline
    \texttt{45med} & $128\times128\times256$ 
          & $45^\circ\times45^\circ$ 
          & $4200\,\mathrm{s}$ \\
    \texttt{45high} & $256\times256\times512$ 
          & $45^\circ\times45^\circ$ 
          & $1100\,\mathrm{s}$ \\
    \texttt{90low} & $128\times128\times128$ 
          & $90^\circ\times90^\circ$ 
          & $2500\,\mathrm{s}$ \\
    \texttt{90med} & $256\times256\times256$ 
          & $90^\circ\times90^\circ$ 
          & $2500\,\mathrm{s}$ \\
    \hline
    \end{tabular}
    \caption{\normalsize 
    Three-dimensional silicon-burning simulations presented in this paper. 
    The resolution is given as 
    $N_\theta\times N_\phi\times N_r$. The final column gives the total 
    physical time evolved in each simulation.}
    \label{tab:Models}
\end{table}


\section{Long term evolution: 1D vs 3D}
\label{subsec:1D_3D}

We begin by comparing the global evolution of our three-dimensional hydrodynamic simulation with the one-dimensional MESA model from which the initial conditions were extracted. We choose to compare the \texttt{45med} simulation, which has run to $\approx$ 4200\,s, corresponding closely to the end of the 1D model. This comparison is \emph{not} intended to be one-to-one. The MESA calculation was evolved using the 22-isotope \texttt{‘approx21\_plus\_Co56’} network, whereas the hydrodynamic simulation uses the 25-isotope network described in Section~\ref{sec:setup}. The 22-isotope network is designed as an efficient reduced network for advanced stellar evolution calculations, including implicit reactions, an approximate treatment of weak interactions and deleptonization approaching core collapse, but it does not contain the same isotopic coverage or reaction pathways as our hydrodynamic network, while the 25-isotope network we use is purely explicit. \textsc{MESA} itself solves the stellar structure and composition equations in one dimension using time-dependent mixing and nuclear burning prescriptions, whereas the three-dimensional calculation evolves the turbulent flow explicitly. These differences make a strictly quantitative comparison of the detailed abundance evolution difficult. Instead, our goal in this section is to compare the broad properties and qualitative behaviours of the silicon-burning shell in the two calculations, and better understand the impact of silicon burning with realistic turbulent mixing. Given our goals, our choice of nuclear network is sufficient. The impact of the nuclear network choice in 3D simulations will be explored in a future study; here we focus on the hydrodynamic consequences of three-dimensional mixing using a fixed 25-isotope network.

\begin{figure*}
\centering
    \subfloat{\includegraphics[width=0.51\linewidth]{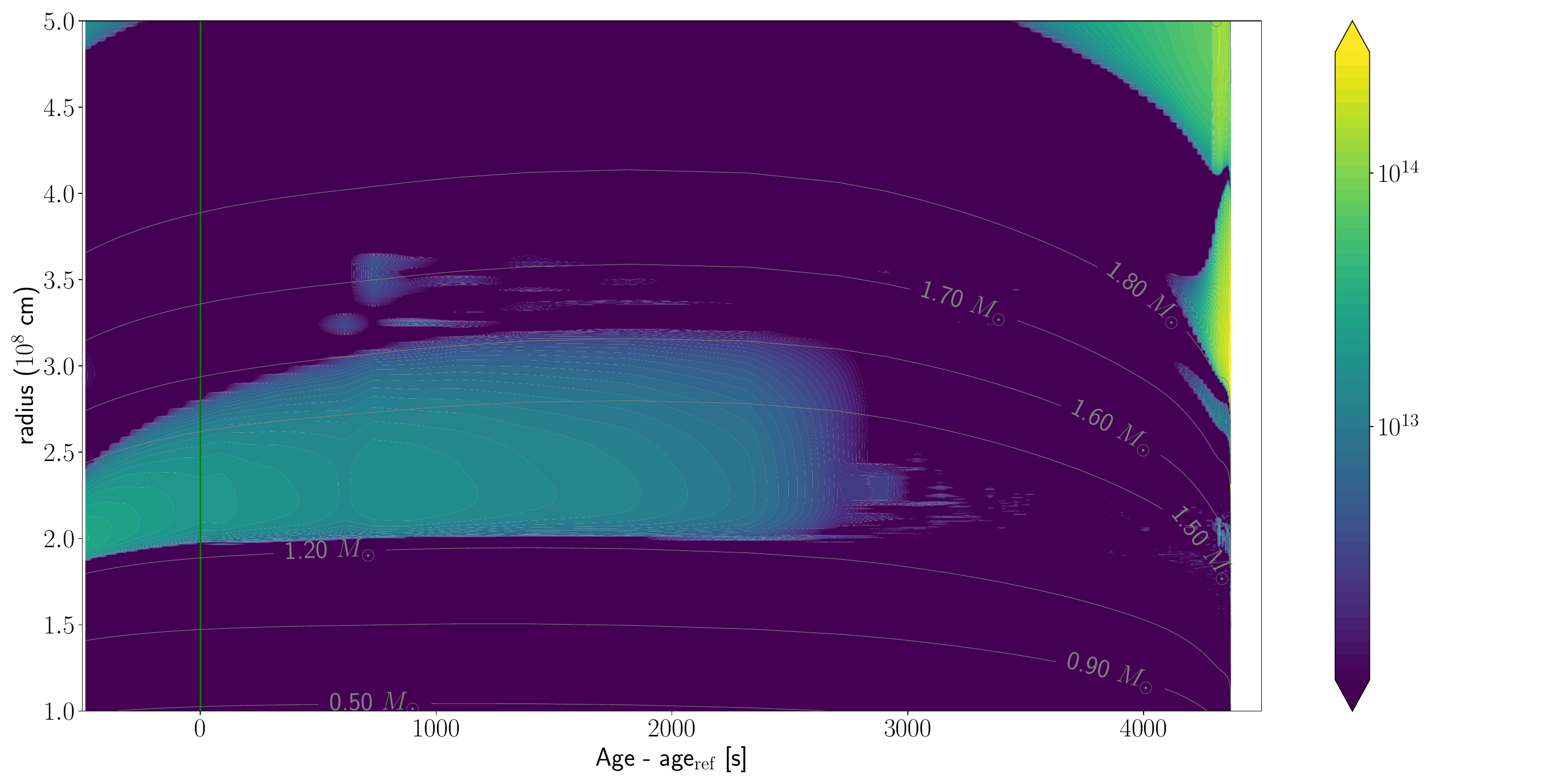}}
\hfil
    \subfloat{\includegraphics[width=0.49\linewidth]{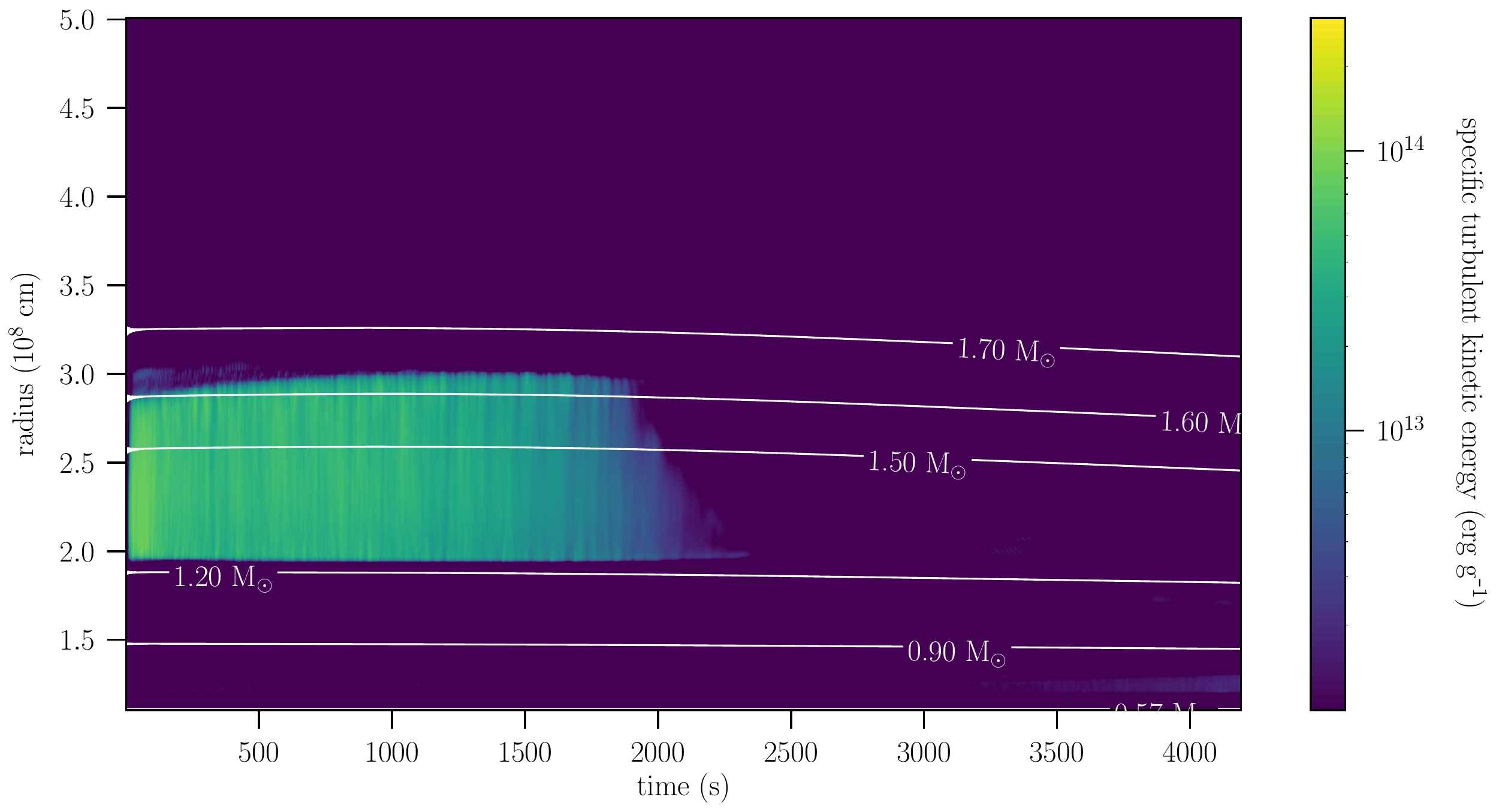}}

\caption{\normalsize Time evolution, in seconds, of the turbulent kinetic energy against radius of the 1D model (left) compared to the 3D model (right). For the 1D model, the time 0, marked with a vertical green line, is the initial condition for the 3D simulation. Both plots include white/gray lines that indicate iso-mass contours.}
    \label{fig:2Dmaps_rad}
\end{figure*}

\begin{figure*}
\centering
    \subfloat{\includegraphics[width=0.51\linewidth]{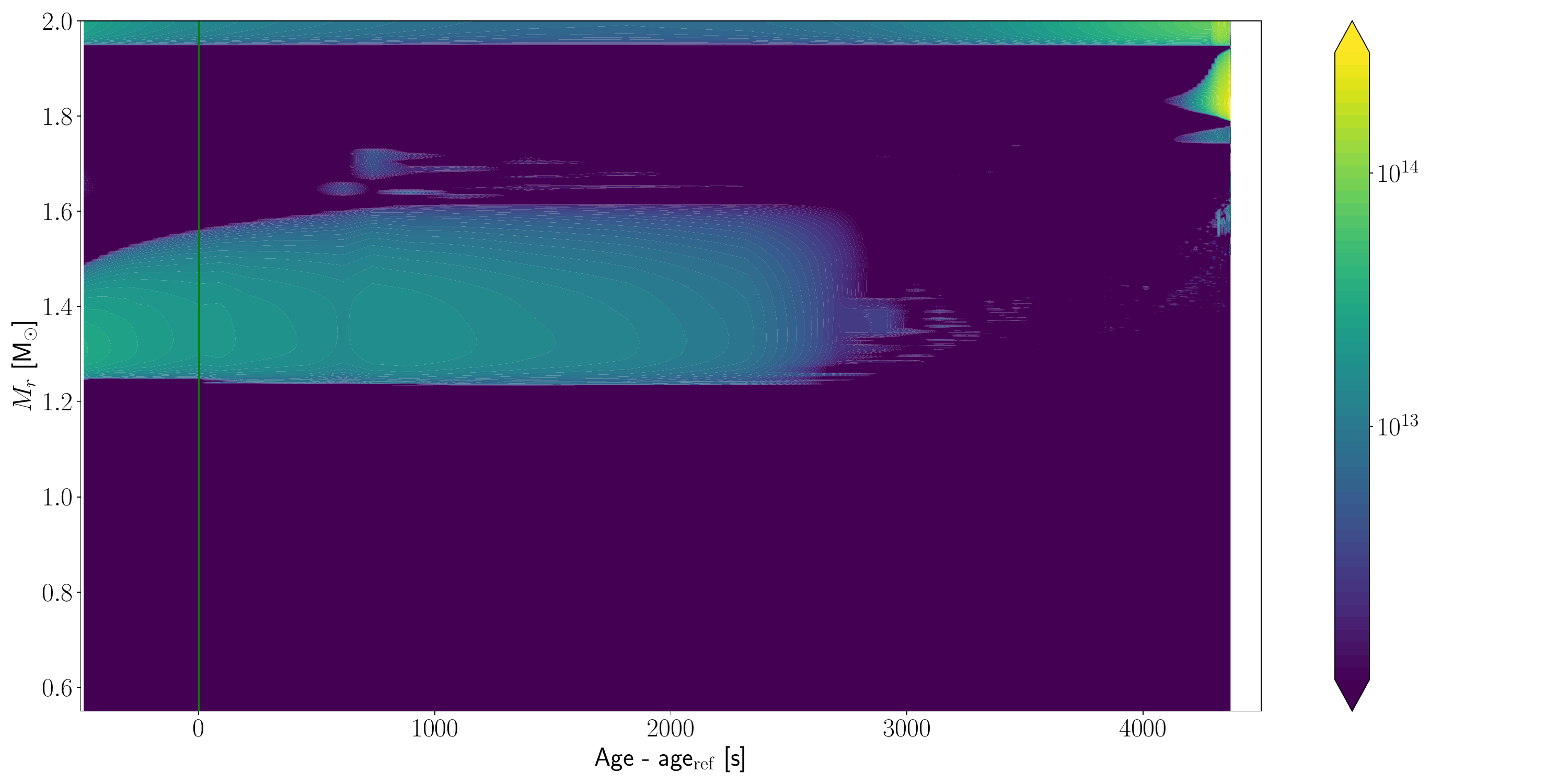}}
\hfil
    \subfloat{\includegraphics[width=0.49\linewidth]{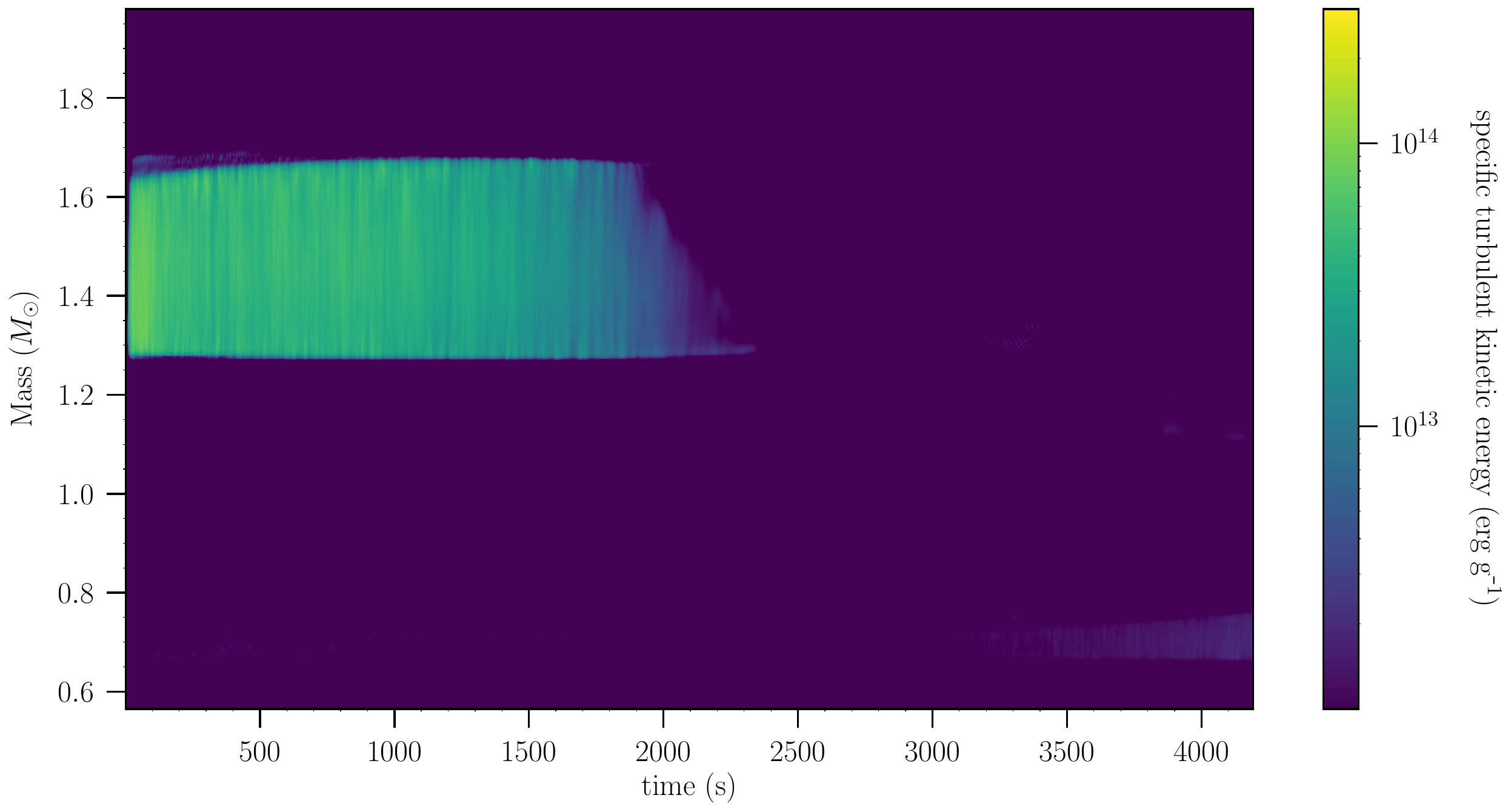}}
    
\caption{\normalsize Time evolution, in seconds, of the turbulent kinetic energy as a function of mass coordinates for the 1D model (left) compared to the 3D model (right). For the 1D model, the time 0, marked with a vertical green line, is the initial condition for the 3D simulation.}
    \label{fig:2Dmaps_mass}
\end{figure*}

Figures~\ref{fig:2Dmaps_rad} and~\ref{fig:2Dmaps_mass} show the time evolution of the specific turbulent kinetic energy in the one-dimensional MESA model compared to the 3D hydrodynamic simulations. In Figure~\ref{fig:2Dmaps_rad}, the vertical coordinate is the radius, while in Figure~\ref{fig:2Dmaps_mass} the same evolution is shown in enclosed mass. In both figures, the time coordinate is measured relative to the moment at which the MESA model was mapped to the three-dimensional hydrodynamic domain. The vertical line at t\,=\,0 therefore marks the beginning of the 3D calculation. This representation allows us to compare the subsequent evolution of the silicon-burning convective region between the original stellar evolution model and the multidimensional simulation. The 3D model we present in both these plots is the \texttt{45med} model, described above.

Comparing these models, we find that the specific turbulent kinetic energy is systematically lower in the \textsc{MESA} model than in the 3D simulation by about a factor of 10. This difference is expected, since previous 3D hydrodynamic simulations have systematically shown that mixing length theory (MLT; \citet{Boehm1958}) used for the one-dimensional convection model tends to under-predict convective velocities. We therefore emphasise the relative timing and radial extent of the burning shell rather than the absolute normalisation of the turbulent kinetic energy. 

After mapping to 3D, the silicon-burning shell evolves in a broadly similar manner to the \textsc{MESA} model, however, it exhausts its fuel earlier than the full one-dimensional calculation. In the \textsc{MESA} model, the silicon burning shell is exhausted after around 3000\,s, while in 3D, it exhausts just after 2000\,s. The longer-lived burning seen in the MESA model arises because fresh silicon-rich material is mixed into the burning region from a previously convective layer above the active silicon shell. We see the evidence of this from around 800\,s in Figure~\ref{fig:2Dmaps_rad} and ~\ref{fig:2Dmaps_mass}, in the regions above $3.0\times10^{8}\mathrm{cm}$ and 1.6\,M$_\odot$ respectively. Here, we see regions above the convective shell that are also becoming convectively unstable, which is not the case for the 3D simulation. 

While these regions seem disconnected from the main convective shell in the TKE plots, this is because their kinetic energy remains very low. However, during the evolution, this entire region above the main convective shell is tagged as being convectively unstable. We checked that this unstable region above the main convective shell was a previously convectively unstable shell. We can see this previously active convective shell in the Kippenhahn diagram in Figure~\ref{fig:init}, and also in the initial entropy profile. Above the silicon shell we are exploring, we find a region with a flat entropy profile, which is a by-product of this earlier convective shell. Once the CBM in the 1D stellar evolution model reaches this flat entropy region, the two shells effectively merge, bringing fresh $^{28}\mathrm{Si}$ and $^{32}\mathrm{S}$ to the silicon shell, extending its life. In contrast, the 3D simulation does not entrain this overlying layer efficiently. Thus, unlike previous studies of convective shells \citep{cristini_17,rizzuti_23, Georgy2024} in this case, the one-dimensional model appears to exhibits more effective CBM than the three-dimensional calculation.

We attribute this difference to how convective boundary mixing is treated in  the 1D model. In this \textsc{MESA} model, mixing beyond the convective boundary is imposed through a diffusive prescription, with an exponentially decaying diffusion coefficient over a length scale set by a fraction of the pressure scale height \citep{Freytag1996, Herwig2000}. Such exponential overshoot prescriptions are widely used in stellar evolution calculations, but they do not explicitly respond to the detailed multidimensional boundary dynamics, or the stiffness of the stable layer beyond the boundary in the same way as a hydrodynamic calculation. Once the diffusive mixing region connects the active silicon shell to the formerly convective layer above it, the flat entropy profile allows this region to mix efficiently into the mixed zone. This artificially expands the effective silicon-burning shell and supplies additional fuel. 

In the hydrodynamic simulation, by contrast, the entropy jump at the upper boundary remains sufficiently large that only limited entrainment occurs. Fresh fuel is still mixed into the silicon-burning shell, but the rate and extent of the CBM region is naturally limited. The shell therefore dies once its fuel is consumed, without entraining much of the silicon-rich material above it.

\begin{figure*}
    \includegraphics[width=\linewidth]{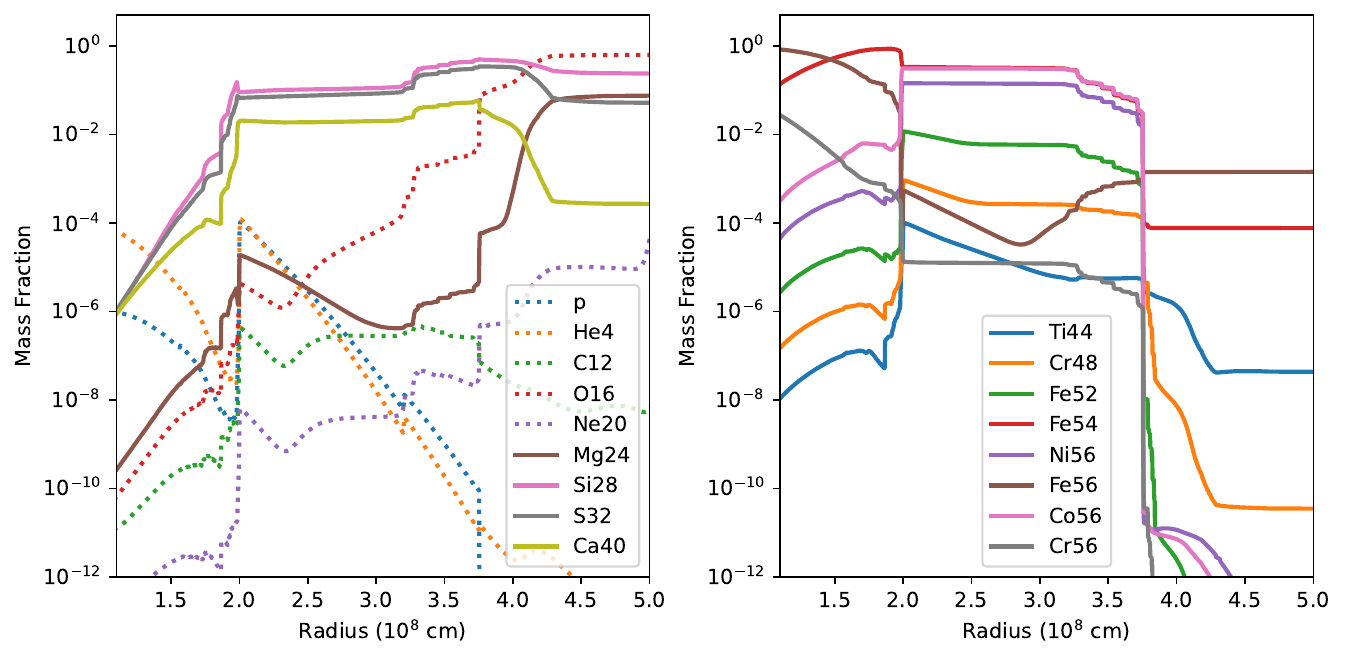}
    \caption{\normalsize  The full range of isotopes in MESA at 1000\,s after the 3D simulation has begun}
    \label{fig:MESA_iso_full}  
\end{figure*}

\begin{figure*}
    \includegraphics[width=\linewidth]{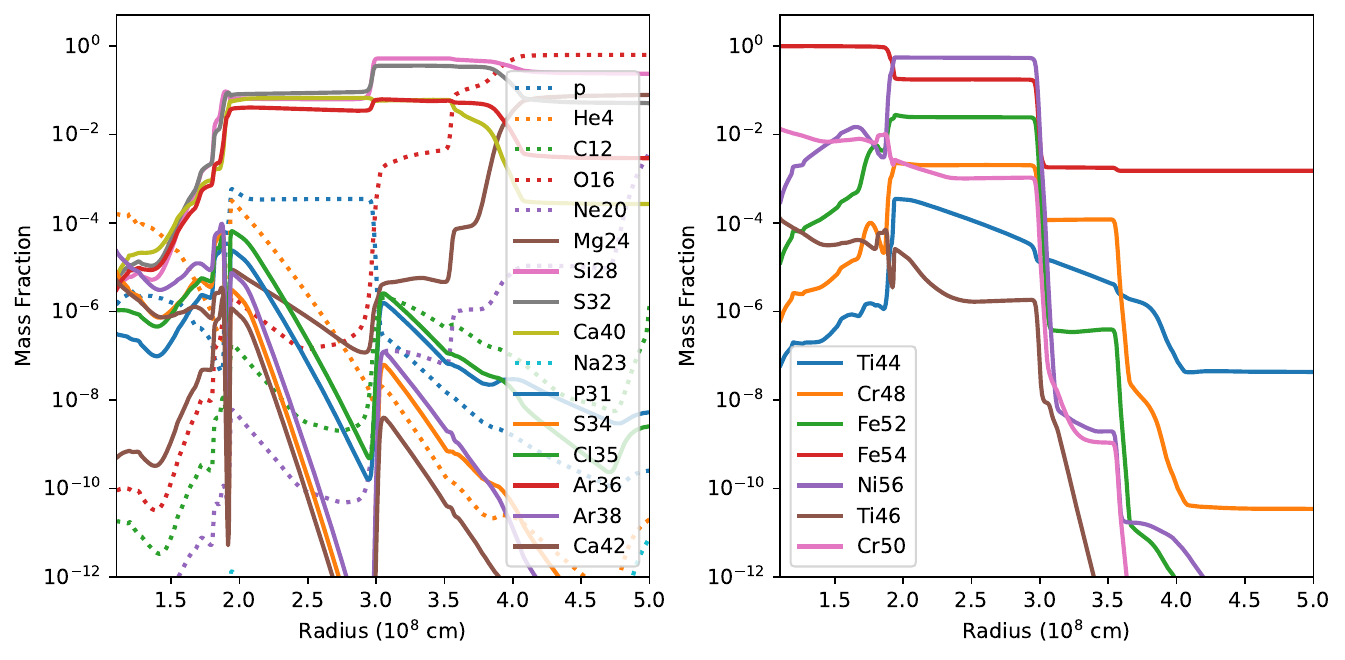}
    \caption{\normalsize  The full range of isotopes in PROMPI at 1000\,s after the 3D simulation has begun}
    \label{fig:PROMPI_iso_full}  
\end{figure*}

We see the immediate consequence of this difference in mixing  in Figures~\ref{fig:MESA_iso_full} and~\ref{fig:PROMPI_iso_full}. These figures compare radial abundance profiles for the full list of isotopes of the MESA and \textsc{PROMPI} models, respectively, at around 1000\,s after mapping. This time corresponds approximately to the phase when the MESA model has already mixed the silicon-burning shell with material out to r$\approx3.6\times10^8$\,cm. In contrast with the 1D model, it is clear that the 3D hydrodynamic simulation maintains sharp abundance and entropy profiles at the upper convective boundary near r$\approx3\times10^8$\,cm, preventing the same large-scale incorporation of the overlying layer. The abundance profiles therefore provide an independent indication that the late-time extension of burning in the MESA model is linked to enhanced one-dimensional boundary mixing. 

Within the convective region itself, both models show significant radial gradients in several isotopes. This is an important signature of convective-reactive burning, where the nuclear burning timescale is comparable to (or smaller than) the mixing timescale, so the composition is not homogenized. In the 3D model, silicon-rich material remains abundant above the active shell, but the burning time at the base of the shell is short enough that entrainment cannot replenish the fuel as rapidly as it is consumed. The convective luminosity and turbulent kinetic energy therefore decline as the burning layer exhausts itself. We note that despite the implicit time-stepping in MESA, it is still able to recover these gradients since mixing and burning are done as a coupled set of equations, which is not always the case in 1D stellar evolution models.

A further difference between the two models is visible in the dominant products of silicon burning. In the MESA model, the final iron-group material is dominated by $^{54}$Fe and $^{56}$Co, whereas our 3D calculation produces a larger fraction of $^{56}$Ni. This difference is largely due to the different nuclear networks and electron-fraction evolution. Silicon burning proceeds through quasi-statistical equilibrium groups, and the final iron-group composition is highly sensitive to the neutron excess \citep{Hix1996}. Material with electron fractions ($Y_e$) close to 0.5 preferentially produces symmetric nuclei such as $^{56}$Ni, while burning happening at lower $Y_e$ favours more neutron-rich species. Although our nuclear reaction network in 3D contains weak reactions, once $^{56}$Ni (or $^{52}$Fe) is produced through $\alpha$-captures during silicon burning, the network does not explicitly contain isotopes that these iron-group isotopes can decay into via electron captures. This keeps $Y_e$ in the shell of the 3D model close to 0.5, and $^{56}$Ni the dominant species produced during its burning phase. We therefore interpret the difference in final iron-group composition primarily as a network and weak-interaction effect, rather than as a direct consequence of multidimensional hydrodynamics alone. These differences, while changing the isotopic abundances, shouldn't affect our qualitative comparison between these two models. 

Finally, we note that many of the isotopes showing the strongest radial gradients in our 3D simulation are not located on the main $\alpha$-chain. This is consistent with the idea that the silicon-burning shell is not simply evolving through a single dominant $\alpha$-capture sequence, but through a coupled network in which leakage between QSE groups, neutron excess, and non-$\alpha$-chain reaction pathways influence the abundance evolution. These gradients provide further evidence that the shell is in a convective-reactive regime, where turbulent transport and nuclear processing must be considered together. We will discuss the nuclear burning that occurs in our 3D model in more detail in Section~\ref{subsec:enuc}.

\section{Resolution and opening angle dependence}
\label{subsec:resolution}

\begin{figure*}
\centering
    \begin{subfigure}{0.4\linewidth}
        \includegraphics[width=\linewidth]{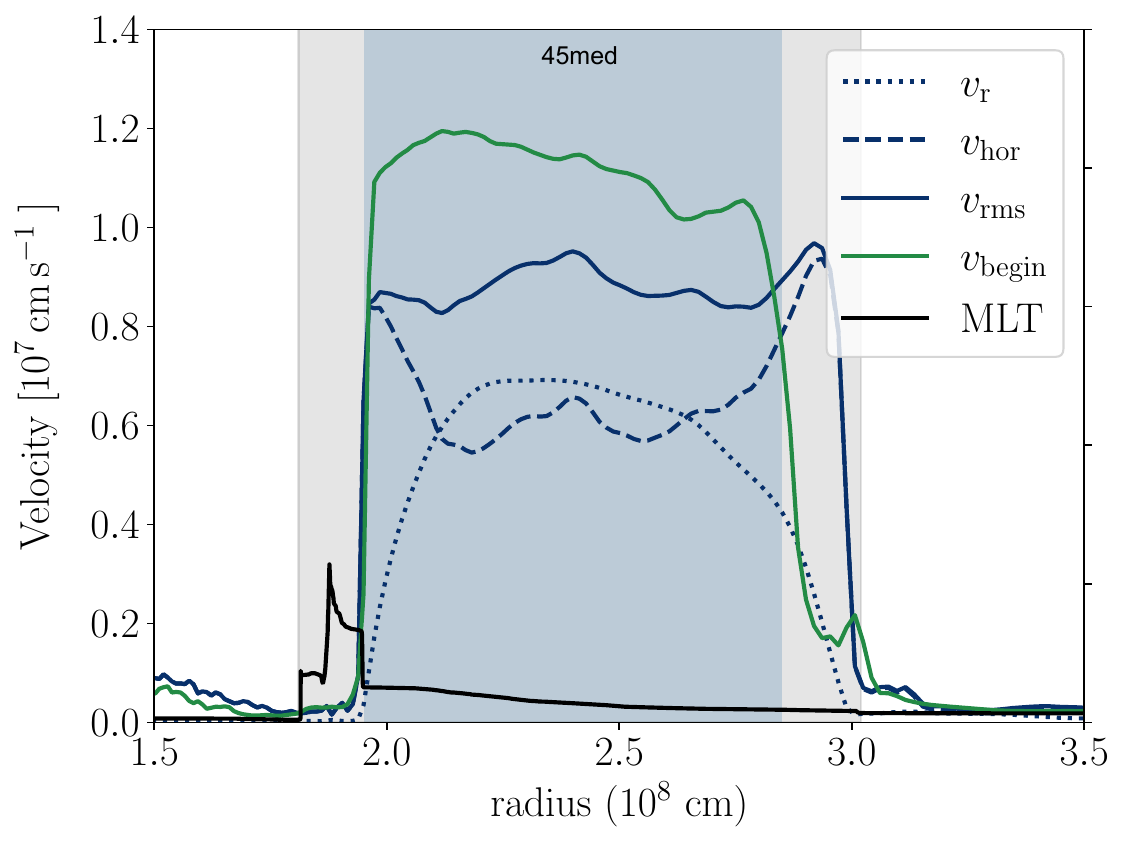}
    \end{subfigure}
\hfil
    \begin{subfigure}{0.4\linewidth}
        \includegraphics[width=\linewidth]{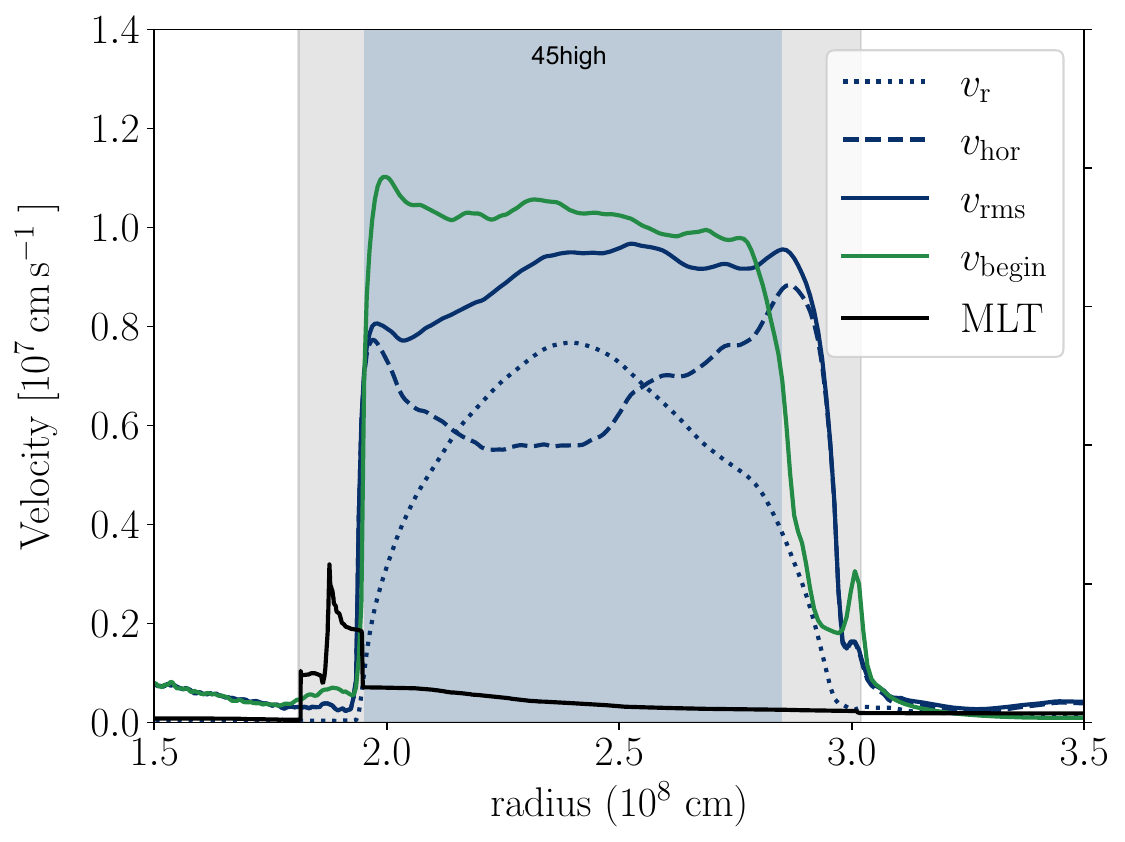}
    \end{subfigure}

    \begin{subfigure}{0.4\linewidth}
        \includegraphics[width=\linewidth]{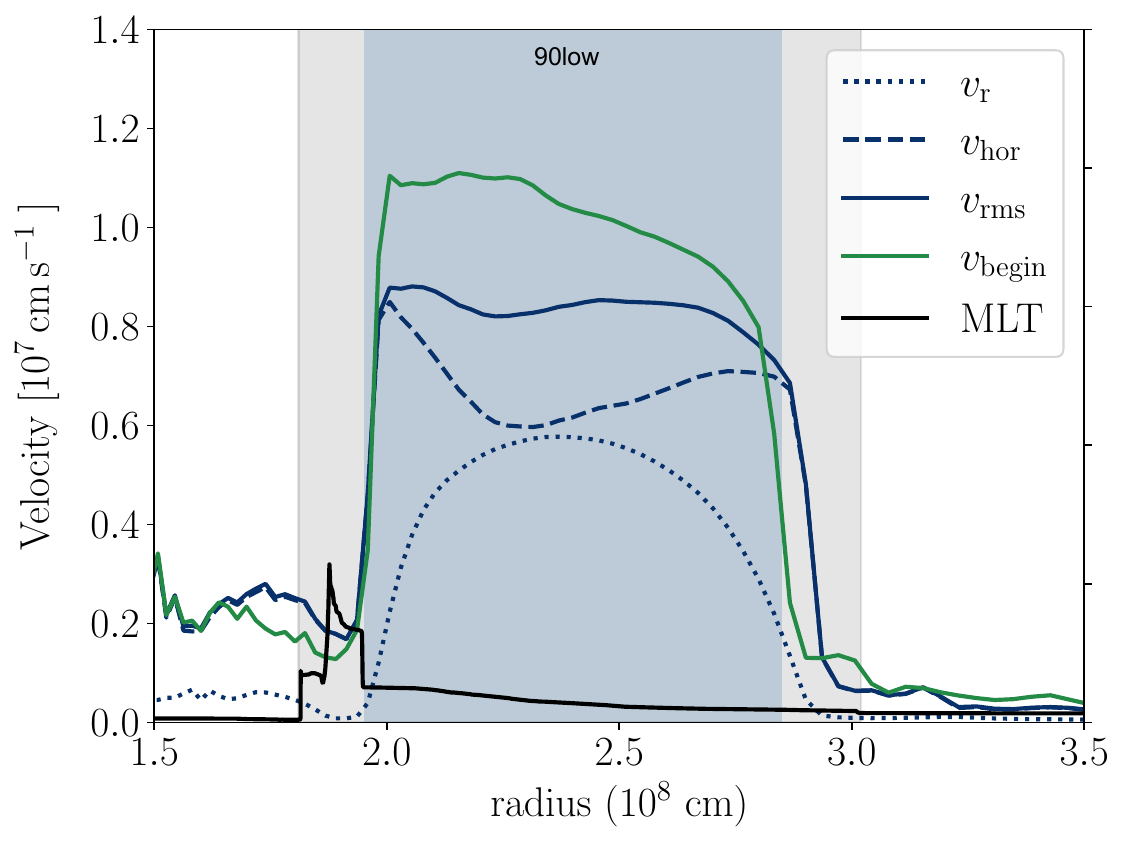}
    \end{subfigure}
\hfil
    \begin{subfigure}{0.4\linewidth}
        \includegraphics[width=\linewidth]{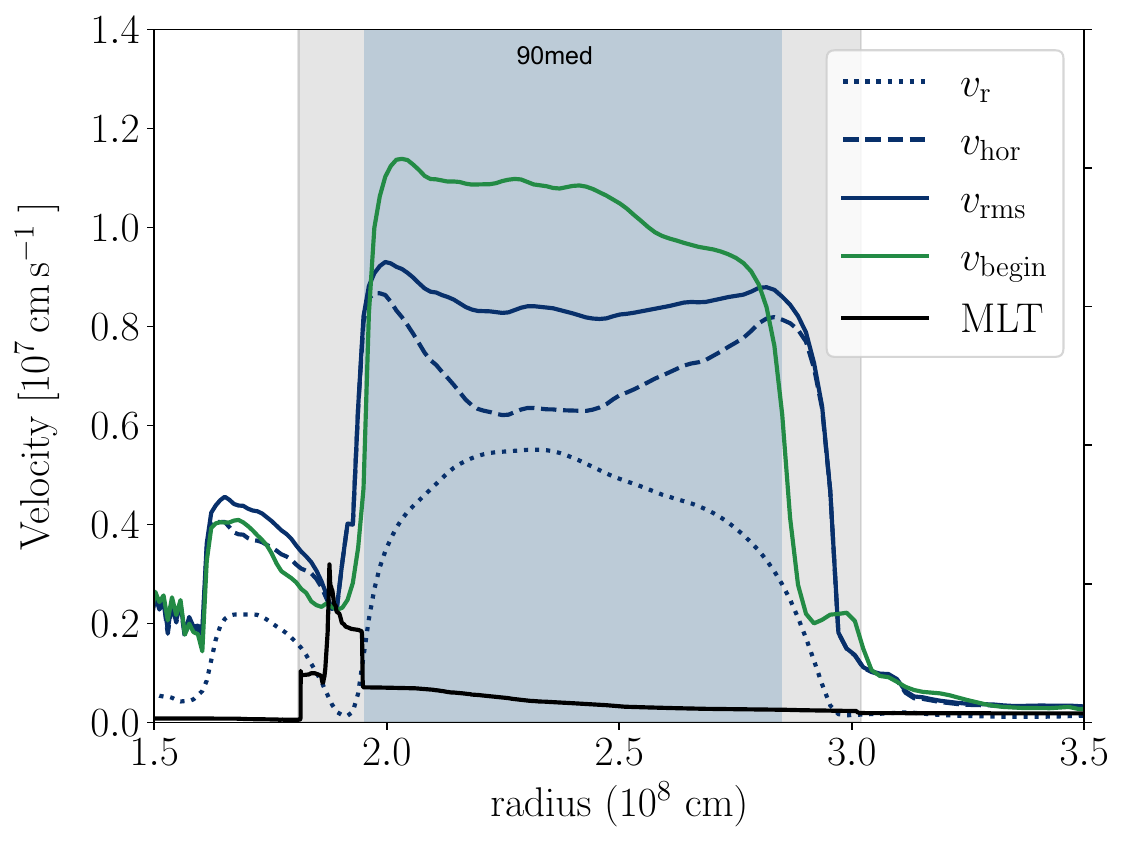}
    \end{subfigure}

\caption{\normalsize 
Velocity profiles for the four simulations. The quantity $v_{\rm begin}$ is 
the rms velocity at $t\simeq100\,\mathrm{s}$, shortly after the initial 
relaxation phase. The profiles $v_r$, $v_{\rm hor}$, and $v_{\rm rms}$ show the 
radial, horizontal, and total rms velocities at $t\simeq1000\,\mathrm{s}$. The 
blue shaded region indicates the initial convective shell inferred from the 
mapped entropy and composition profiles, while the grey shaded region indicates 
the region with non-zero MLT velocity in the original MESA model. }
    \label{fig:Velocity_profile}
\end{figure*}

\begin{figure*}
\centering
    \begin{subfigure}{0.4\linewidth}
        \includegraphics[width=\linewidth]{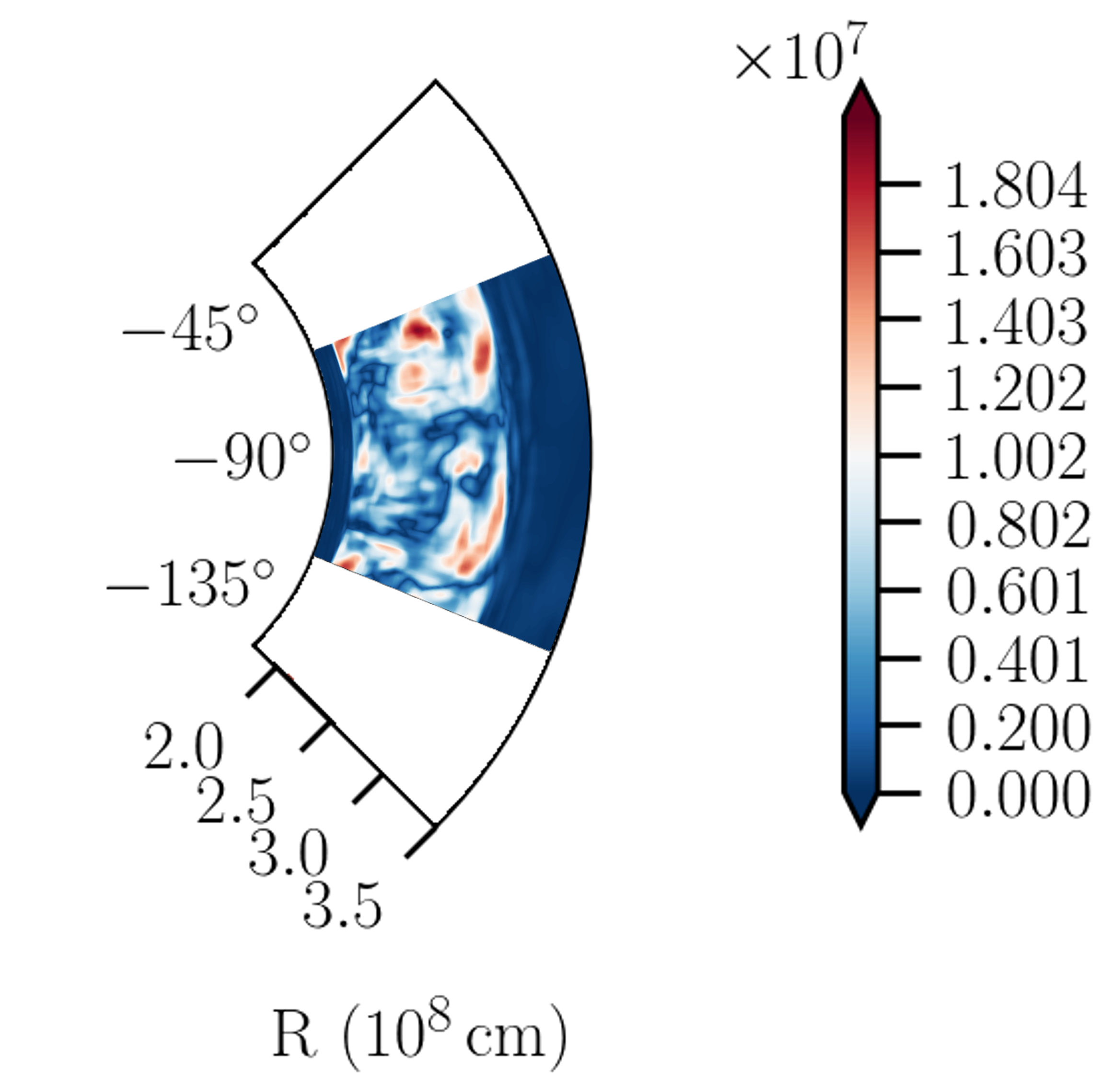}
    \end{subfigure}
\hfil
    \begin{subfigure}{0.4\linewidth}
        \includegraphics[width=\linewidth]{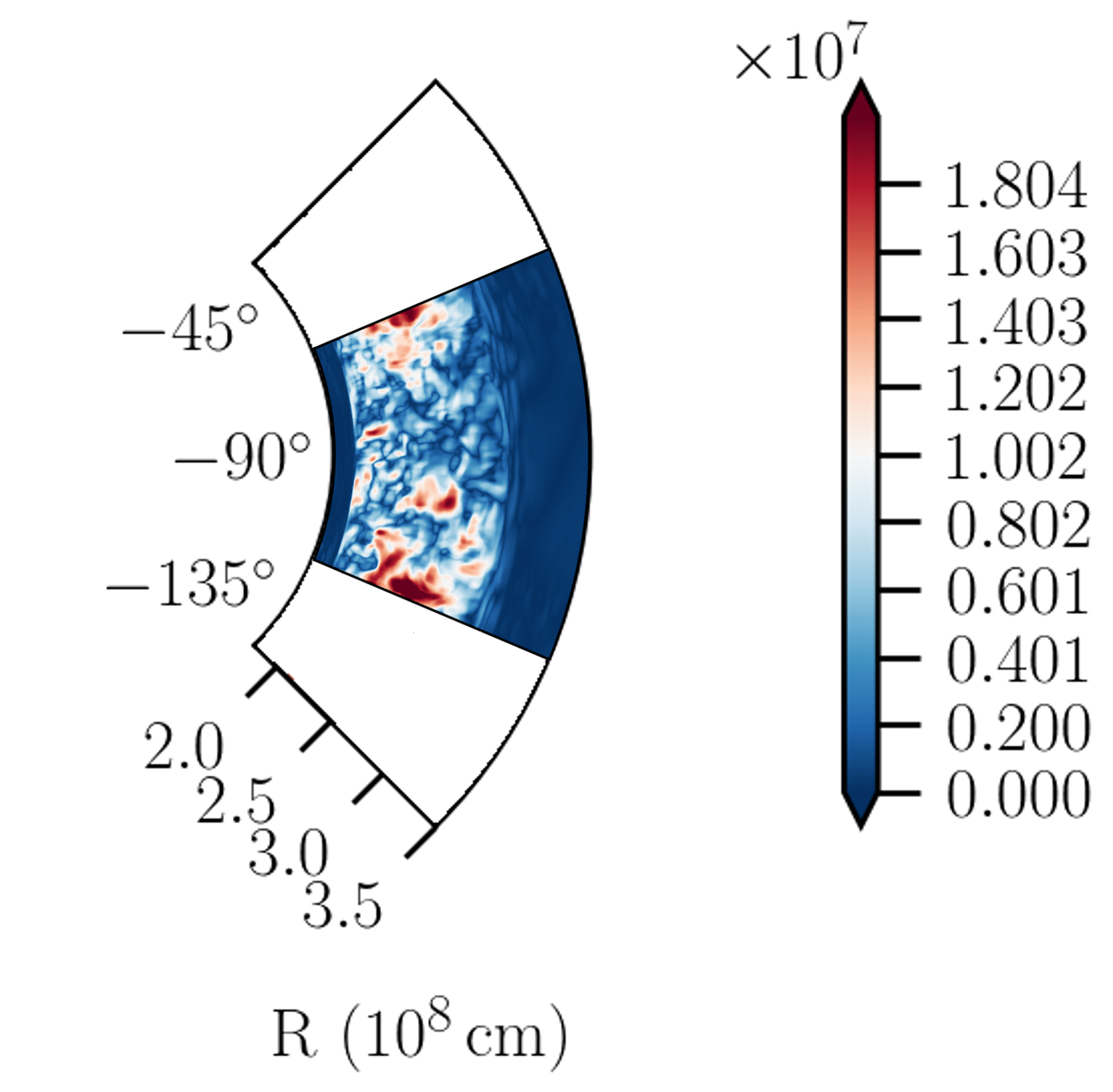}
    \end{subfigure}

    \begin{subfigure}{0.4\linewidth}
        \includegraphics[width=\linewidth]{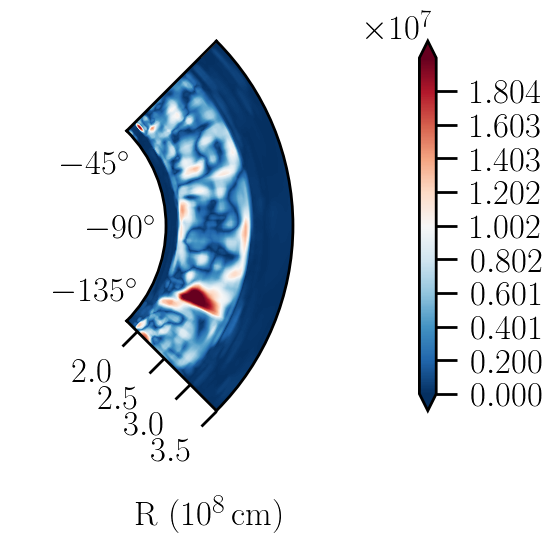}
    \end{subfigure}
\hfil
    \begin{subfigure}{0.4\linewidth}
        \includegraphics[width=\linewidth]{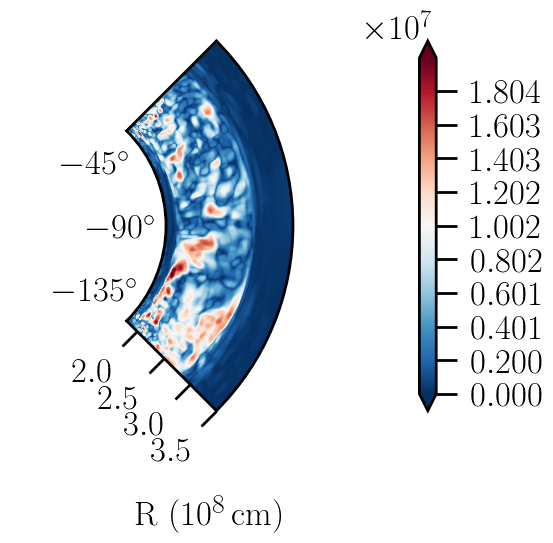}
    \end{subfigure}

\caption{\normalsize 
Two-dimensional slices of the velocity magnitude for the four simulations at 
$t\simeq1000\,\mathrm{s}$. The panels compare the effect of both angular 
opening angle and numerical resolution. From left to right, top to bottom, the models visualised are \texttt{45med}, \texttt{45high}, \texttt{90low} and \texttt{90med}.}
    \label{fig:2D_Slice}
\end{figure*}

In this section, we assess the sensitivity of the silicon-shell convection 
to angular extent and grid resolution. Figure~\ref{fig:Velocity_profile} compares the 
corresponding spherically averaged velocity profiles for all our models while Figure~\ref{fig:2D_Slice} shows 
two-dimensional slices of the velocity magnitude for the four simulations at 
$t\simeq1000\,\mathrm{s}$. We show the rms velocity 
soon after the initial transient, at $t\simeq100\,\mathrm{s}$, denoted 
$v_{\mathrm{begin}}$, together with the radial velocity, horizontal velocity, and 
total rms velocity at $t\simeq1000\,\mathrm{s}$. For comparison, we also show 
the convective velocity predicted by MLT in the original MESA model at the time 
of mapping.

The blue shaded region in Figure~\ref{fig:Velocity_profile} marks the initial 
convective shell inferred from the entropy and composition profiles of the 
mapped model. This region corresponds well to the location where convection 
develops in the early 3D velocity profiles. The broader grey shaded region indicates 
the region with non-zero MLT velocity in the MESA model. The MESA profile is based 
on a the MLT convection prescription and on the nuclear network used in 
the stellar evolution calculation, whereas the hydrodynamic simulations use a 
different reduced network and evolve the velocity field explicitly. Small 
differences in the reaction network, especially in the reverse rates important 
for silicon burning, can shift the location of the peak energy generation and hence 
the inferred convective driving region. 
In addition, the MLT velocity is a local 
one-dimensional estimate and is not expected to reproduce the detailed shape of 
the resolved 3D velocity profile.

Comparing the four hydrodynamic simulations, we find that the early time
rms velocity profiles, $v_{\mathrm{begin}}$, are broadly similar. In all models, $v_{\rm begin}$ peaks 
at approximately $v_{\rm rms}\simeq(1.0-1.1)\times10^7\,\mathrm{cm\,s^{-1}}$
and the radial extent of the convective shell is comparable. This suggests that 
the bulk velocity scale of the silicon-burning convection is reasonably robust 
to the range of resolutions and angular opening angles considered here. 
At $t\simeq1000\,\mathrm{s}$, the total rms velocity profiles have all dropped in their magnitude to $\simeq 0.8\times10^7\,\mathrm{cm\,s^{-1}}$ but remain qualitatively 
similar between the models. There are, however, differences in the 
precise location and sharpness of the upper convective boundary. The \texttt{45med} model in particular has the most extended upper convective boundary. It is unclear from just these 4 models if this is a purely stochastic phenomenon, since model \texttt{90med} has the same local resolution, but does not show a similar growth.

A clearer difference appears when the radial and horizontal velocity components 
are considered separately. In the $45^\circ$ models, the radial and horizontal 
velocity profiles have the expected shell convection structure: radial motions 
are strongest in the interior of the convective zone, while the horizontal 
velocity becomes more important near the convective boundaries, 
where rising and sinking plumes turn over. This behaviour is consistent with 
previous multidimensional simulations of deep stellar convection, including 
recent neon-shell calculations, where the horizontal velocity peaks near the 
boundaries because of plume deflection and shear at the stable interfaces 
\citep{Georgy2024}. 

The $90^\circ$ simulations show the same broad convective region but exhibit a 
more pronounced horizontal velocity component compared to their radial velocities. We interpret this as a possible 
geometrical effect associated with the thinness of the silicon-burning shell 
and the stiffness of its convective boundaries. In a radially thin shell bounded 
by strong buoyancy barriers, radial plume penetration is inhibited, and the 
flow may be redirected preferentially into horizontal motion. Similar behaviour 
has been observed in the thin silicon-burning shells in
\citet{Griffiths2026}. In our models, this effect is 
most visible in the larger angular domains, because the $90^\circ$ simulations 
allow larger horizontal flow structures to develop than the $45^\circ$ wedges. 
Whether this altered flow morphology has a significant impact on convective 
boundary mixing or fuel entrainment remains to be confirmed by future studies.

A more important difference between the two angular domains appears below the 
main convective shell, at radii 
$r\lesssim1.8\times10^8\,\mathrm{cm}$. In this region, the $90^\circ$ models 
show noticeably larger velocities than the $45^\circ$ models. These motions do 
not appear to be part of the physical silicon-burning convection zone. Instead, 
we attribute them to high-frequency numerical artefacts associated with the 
larger angular extent of the spherical polar grid and its proximity to the 
coordinate singularity. The problem is exacerbated by the sensitivity of 
silicon burning to small thermodynamic and compositional perturbations. Even 
weak numerical velocity fluctuations below the convective boundary can induce 
mixing, alter the local composition, and trigger additional burning.
To reduce these artefacts, we have applied a fourth-order Shapiro filter \citep{Shapiro1970} in the 
angular directions for $\theta>60^\circ$. The filter is 
applied only below the convective zone and is designed to remove short-wavelength 
oscillations without modifying the large-scale convective flow. This procedure 
substantially reduces the numerical noise, but it does not remove it completely. 
Over time, the residual fluctuations still mix material below the lower 
convective boundary and can sustain artificial burning and enhanced velocities 
in this region. These artefacts can also be seen visually at the boundaries of the velocity slices we present in Figure~\ref{fig:2D_Slice}. We note that they are also less pronounced in \texttt{90low} compared to \texttt{90med} since more of these small scale fluctuations are diffused away due to the higher numerical diffusivity in this model.

For this reason, we regard the $90^\circ$ models as useful tests of angular 
opening angle and large-scale flow morphology, but not as our most reliable 
models for detailed quantitative analysis of convective boundary mixing or 
nucleosynthesis. The bulk properties of the main convective shell are similar 
across all four simulations, which gives confidence that the primary conclusions 
are not strongly resolution-dependent. However, because the $45^\circ$ models 
are less affected by the low-radius numerical artefacts, we focus most of the 
subsequent analysis on the two $45^\circ$ simulations, using the higher-resolution 
model when examining quantities that are likely to be resolution-sensitive.

\section{Structure and evolution of the velocity field and boundary mixing}
\label{subsec:velocity_field}

\begin{figure}
 \includegraphics[width=\linewidth]{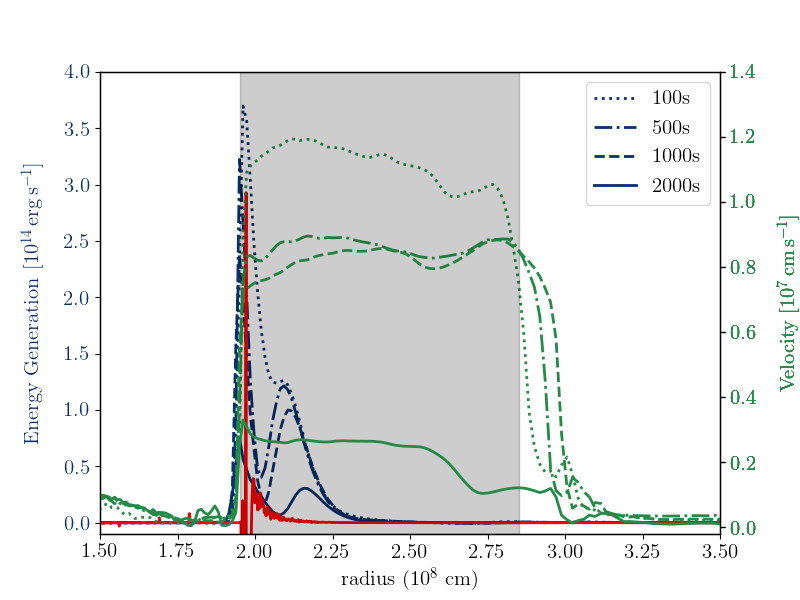}
 \caption{\normalsize Time evolution of energy generation and radial RMS velocity profiles. The red line shows the energy generation rate from the 1D MESA model for comparison.}
 \label{fig:enuc_profiles}  
\end{figure}

In Figure~\ref{fig:enuc_profiles}, we show the time evolution of the horizontally averaged radial velocity and the nuclear energy-generation rate of model \texttt{45med}. The early evolution is dominated by a short adjustment phase following the mapping of the one-dimensional progenitor to the three-dimensional hydrodynamic grid. During the first $\approx 200\,{\rm s}$, the radial velocities are enhanced as the model relaxes away from the mapped initial condition. After this initial transient, the convective shell settles into a quasi-steady state for most of the active silicon-burning phase. During this period, the mean radial velocity remains approximately constant, consistent with the largely steady turbulent kinetic energy seen in Figures~\ref{fig:2Dmaps_rad} and \ref{fig:2Dmaps_mass}. The main variation we see in the radial velocity profile, is after 2000\,s, which corresponds to when the shell is running out of fuel, and hence the radial velocity drops substantially. From Figure~\ref{fig:2Dmaps_rad} and ~\ref{fig:2Dmaps_mass}, it is clear that this eventually drops to zero.

The energy-generation profile shows a qualitative change in its structure over time. In particular, a double-peaked structure develops in $\epsilon_{\rm nuc}$ after the initial transient. We note also that in the \textsc{MESA} model the structure and position of the energy generation profile are transient. While in the snapshot we show, the profile looks qualitatively similar to the 3D model, at even slightly earlier or later times, the position and magnitude of these peaks change quite drastically. The inner-most positive peak sometimes moves further radially inwards, explaining the region that is Schwarzschild unstable marked in Figure~\ref{fig:Velocity_profile}. This is largely due to the fact that many of the forward and reverse reactions along the $\alpha$ capture chain have very similar reaction rates. So slight variations in the temperature, density and compositions at each timestep can qualitatively alter the energy generation profile.
We return to the origin of this double-peaked feature in Section~\ref{subsec:enuc}. Here, we note that the changing radial distribution of nuclear burning does not appear to produce a qualitative change in the large-scale flow morphology. Despite this change in the burning profile, the convective flow remains dominated by broad, shell-scale upflows and downflows, and we find no evidence for a major reorganization of the velocity field.

Although some CBM is present, the net entrainment of material into the silicon-burning convection zone is modest in 3D. This is apparent in Figures~\ref{fig:2Dmaps_rad} and \ref{fig:2Dmaps_mass}. After $\approx 500 \,$ s, both the inner and outer boundaries of the convective shell varies only slightly until the fuel supply is depleted and the shell begins to decay. This behaviour differs from previous multidimensional simulations of late-stage shell convection, where turbulent entrainment produces significant growth of the convective region over time \citep[e.g.][]{Meakin_2007, cristini_17, Rizzuti2022, rizzuti_23, Georgy2024}, especially compared to their 1D counterparts. In the present silicon-burning shell, the convective boundaries are extremely stiff, and the entrainment rate is too small to replenish fuel at the rate required to sustain burning indefinitely.

\begin{figure}
    \includegraphics[width=\linewidth]{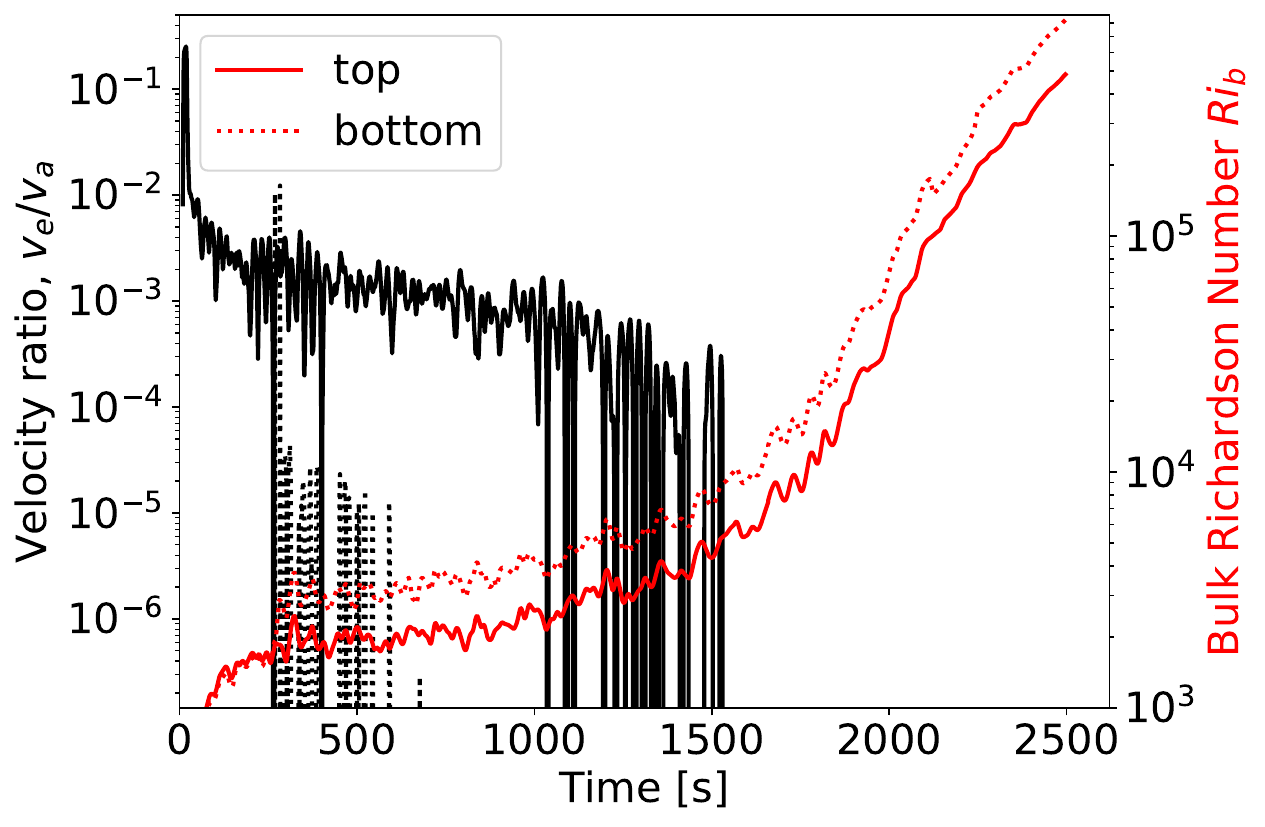}
    \caption{\normalsize  Time evolution plot of the bulk Richardson, $Ri_b$, number (red) and entrainment velocity, $v_e/v_a$, (black). The solid and dotted lines represent the radially outer and inner convective boundaries respectively. }
    \label{fig:RIB}  
\end{figure}

To quantify the stiffness of the convective boundaries, we compute the bulk Richardson number, $Ri_B$, following the entrainment framework introduced in stellar-interior simulations by \citet{Meakin_2007}. The bulk Richardson number compares the stabilizing buoyancy jump across a convective boundary to the kinetic energy of the turbulent flow,
\begin{equation}
    Ri_B = \frac{\ell \Delta b}{v_{\rm rms}^{2}},
    \qquad
    \Delta b = \int_{r_b-\ell/2}^{r_b+\ell/2} N^2 \, dr ,
    \label{eq:rib}
\end{equation}
where $N$ is the Brunt--Väisälä frequency, $r_b$ is the convective boundary radius, $\ell$ is a characteristic turbulent length scale, and $v_{\rm rms}$ is the rms convective velocity. Larger values of $Ri_B$ correspond to stiffer boundaries, for which turbulent eddies have insufficient kinetic energy to efficiently entrain material from the neighbouring stable layer.

The connection between $Ri_B$ and convective boundary mixing is commonly expressed through an entrainment law,
\begin{equation}
    E \equiv \frac{v_e}{v_{\rm rms}} = A Ri_B^{-n},
    \label{eq:entrainment_law}
\end{equation}
where $v_e$ is the entrainment velocity (i.e. the velocity at which the convective boundary moves due to entrainment), and $A$ and $n$ are dimensionless calibration constants. The normalization $A$ sets the entrainment efficiency at $Ri_B=1$, while $n$ determines how rapidly entrainment decreases as the boundary becomes stiffer. Previous stellar hydrodynamic simulations generally find $n$ of order unity, although the precise values of $A$ and $n$ vary between burning phases and progenitors \citep{Meakin_2007, Rizzuti2022, rizzuti_23}.

Figure~\ref{fig:RIB} shows the time evolution of $Ri_B$ and the velocity ratio ${v_e}/{v_{\rm rms}}$ at the inner and outer boundaries of the silicon-burning shell (in dotted and solid lines respectively). Both boundaries have large $Ri_B$ values throughout most of the simulation, indicating that they are dynamically stiff. 
Comparing these values to $Ri_{B}$ values shown in Figure 5 of \citet{Rizzuti2022} and Figure 9 of \citet{rizzuti_23}, it is immediately clear that the $Ri_{B}$ of both boundaries of this silicon burning shell is on the high end compared to other previous shell burning simulations during its quasi-steady burning stage, but both boundaries continue to gradually get stiffer over time. Unsurprisingly, this also corresponds to a gradual decline in the mass entrainment rate, ${v_e}/{v_{\rm rms}}$, over time. Already from early in the simulation, the lower, stiffer boundary has a very low entrainment rate, that effectively goes to zero after 500\,s. ${v_e}/{v_{\rm rms}}$ in the upper boundary also drops to zero after $\approx$ 1500\,s. 

The rise of $Ri_B$ at late times is physically informative. Once the available silicon-rich fuel begins to deplete faster than it is replenished by entrainment, due to the convective-reactive nature of this burning stage, the nuclear energy generation declines. This reduces buoyant driving, which, in turn, reduces convective velocity. Since $Ri_B \propto v_{\rm rms}^{-2}$, the weakening of the turbulent flow causes both boundaries to become stiffer. The increase in $Ri_B$ therefore marks the transition from a quasi-steady burning shell to a decaying convective shell. In this phase, boundary mixing becomes even less effective, accelerating the shutdown of the silicon-burning convection zone. This provides a natural explanation for the limited growth of the convective shell seen in Figures~\ref{fig:2Dmaps_rad} and \ref{fig:2Dmaps_mass}. The TKE available at the boundaries is small compared to the buoyancy barrier that must be overcome to entrain material.

A direct comparison with the one-dimensional model is non-trivial, since the MESA calculation imposes boundary mixing through a diffusive overshoot prescription rather than through a resolved entrainment velocity. Nevertheless, the nearly stationary boundaries in the 3D simulation indicate that the effective hydrodynamic entrainment is much weaker than the mixing implied by the one-dimensional CBM treatment in this model.

We next analyse the flow using the mean-field Reynolds/Favre averaging framework commonly used in Reynolds-averaged implicit large-eddy simulations (RA-ILES) of stellar convection \citep{Chassaing2002, Mocak2014, cristini_17, Georgy2024}. For any quantity $q$, we define the Reynolds decomposition
\begin{equation}
    q = \overline{q} + q',
\end{equation}
and the density-weighted Favre average
\begin{equation}
    \widetilde{q} = \frac{\overline{\rho q}}{\overline{\rho}},
    \qquad
    q = \widetilde{q} + q'' .
\end{equation}
In general, $q'\neq q''$, and the Favre decomposition is preferable for compressible, stratified flows because it separates mean advection from turbulent mass transport. $\overline{q}$ represents spatial and temporal averages of the physical quantity in question.

Here we focus on the radial turbulent kinetic energy budget. We define the radial component of the turbulent kinetic energy as
\begin{equation}
    k^r = \frac{1}{2}\widetilde{u_r''u_r''}.
\end{equation}
The corresponding mean-field equation can be written schematically as
\begin{equation}
   \partial_t(\overline{\rho}\widetilde{k}^r)
   + \nabla_r\left(\overline{\rho}\widetilde{u}_r\widetilde{k}^r\right)
   + \nabla_r f_k^r
   + \nabla_r f_P
   =
   W_B + W_P + W_S + \mathcal{N}_k^r ,
   \label{eq:TKE}
\end{equation}
where $f_k^r$ is the turbulent kinetic energy flux, $f_P$ is the acoustic flux, $W_B$ is the buoyancy work, $W_P$ is the pressure-dilatation work, $W_S$ is the shear-production term, and $\mathcal{N}_k^r$ is the residual. The residual contains the net effect of numerical dissipation and any unclosed terms associated with the finite-volume discretization scheme. In our ILES approach, where no explicit viscosity is included, this term provides an estimate of the effective dissipation occurring near the grid scale.

\begin{figure}
    \includegraphics[width=\linewidth]{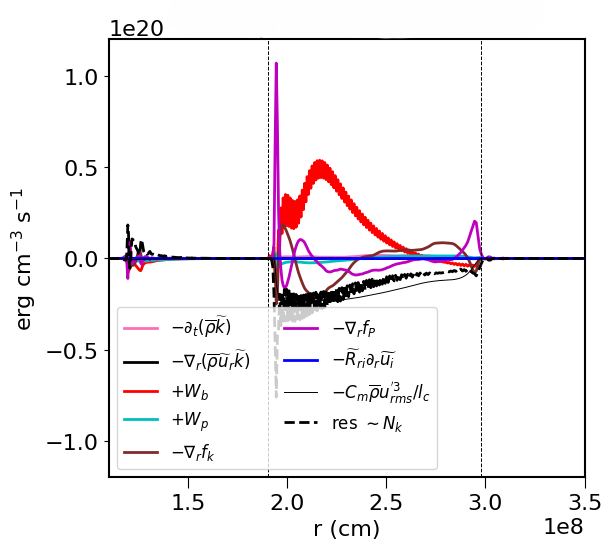}
    \caption{\normalsize Spatial and temporal averaged terms from the turbulent kinetic energy equation. This plot is of the model \texttt{45high} and has been averaged over 10 convective turnovers. $C_m$ here is set to 1.0, showing that the numerical dissipation is well described by isotropic turbulence.}
    \label{fig:TKE}  
\end{figure}

We have displayed each term of the equation in Figure~\ref{fig:TKE}. In particular, this plot depicts our \texttt{45high} model, where we have temporally averaged over three convective turnovers between a physical time of 850\,s and 1000\,s. The vertical dotted lines mark the inner and outer convective boundaries at this time. We note that the Eulerian time derivative, $\partial_t(\Bar{\rho}\Fav{k}^r)$ is negligible in Figure~\ref{fig:TKE}, implying that in our choice of time averaging, the convection is in a statistical steady state. Similarly, the mean advection of TKE ($ \nabla_r(\Rey{\rho}\Fav{u}_{r}\Fav{k}^r)$) and the shear production ($\Fav{R}_{rr}\partial_r\Fav{u}_r$, where $R_{rr}$ is the radial component of the Reynolds stress) are also negligible in our simulation, so we do not discuss these further. Both these terms depend on the gradient of the mean radial velocity which we expect to be small inside a convective region. The TKE budget is therefore controlled primarily by buoyancy driving, pressure work, turbulent and acoustic transport, and numerical dissipation.

The transport terms redistribute the turbulent kinetic energy within the convective shell. The turbulent kinetic energy flux,
\begin{equation}
    f_k^r = \overline{\rho u_r'' k^{r\prime\prime}},
\end{equation}
describes the transport of radial turbulent kinetic energy by correlated velocity fluctuations. The acoustic flux,
\begin{equation}
    f_P = \overline{P' u_r'},
\end{equation}
describes the transport of mechanical energy by pressure fluctuations. As found in previous simulations of deep stellar convection, these two fluxes often partially oppose one another within the convection zone \citep{Viallet2013, cristini_17, Georgy2024}. However, in the present work, near the convective boundaries the acoustic flux becomes more prominent, consistent with the excitation of internal gravity waves in the neighbouring stable layers.

The dominant source of turbulent kinetic energy is buoyancy work,
\begin{equation}
    W_B = \overline{\rho' u_r'}\,g_r ,
\end{equation}
with the sign convention chosen such that positive $W_B$ corresponds to driving of turbulent motions. We find $W_B>0$ throughout most of the convective shell, as expected for buoyantly driven convection. Near the convective boundaries, $W_B$ becomes negative, indicating that radial motions are decelerated as plumes work against stable stratification.

The pressure-dilatation term,
\begin{equation}
    W_P = \overline{P' \nabla\cdot\mathbf{u}'},
\end{equation}
captures the exchange between turbulent kinetic energy and internal energy through compressive motions. In our simulation, $W_B$ shows a less smooth radial structure than in some previous shell-convection calculations. In particular, the region near $r\simeq2\times10^8\,{\rm cm}$ is affected by the irregular double-peaked energy-generation profile discussed above. Although this produces a local structure in the buoyancy driving, the burning region is radially narrow, and the global convective morphology remains largely unchanged.

Finally, the residual term $\mathcal{N}_k^r$ is negative over most of the convective shell, as expected for a net sink of turbulent kinetic energy. Since the simulations do not include an explicit physical viscosity, dissipation occurs through the numerical scheme at the grid scale. The magnitude and radial distribution of $\mathcal{N}_k^r$ therefore provide a useful diagnostic of where TKE cascades out of the resolved flow. Similar to what was found in \citet{Arnett2009}, the numerical dissipation is well described by the local isotropic turbulence expression, $\mathcal{N}_k \approx \Bar{\rho}u'^3_\mathrm{rms}/l_c$, where $l_c$ is the length scale of the convective zone. The approximate balance between buoyancy driving, transport, and residual dissipation supports the interpretation that the shell is in a quasi-steady convective state during the averaging interval. We note that some of the source terms and fluxes show fluctuation below $1.5\times10^8\,\mathrm{cm}$, this is due to gravity waves, launched from the lower boundary of the convection zone interacting with the simulation grid boundary at $1.1\times10^8\,\mathrm{cm}$.

Taken together, these diagnostics indicate that the silicon-burning shell undergoes three phases. First, the mapped model relaxes through a short transient during which the velocities and turbulent kinetic energy are enhanced. Second, the shell enters a quasi-steady convective-burning phase in which buoyancy driving maintains a roughly constant velocity scale, while the convective boundaries remain stiff and only weakly entrained. Finally, as silicon-rich fuel is consumed faster than it can be replenished by boundary mixing, the nuclear energy generation rate declines, leading to a decrease in the convective velocity and hence a rise in the bulk Richardson number, $Ri_{b}$. The increasing boundary stiffness then further suppresses entrainment, until the shell exhausts its available fuel. In comparison, the 1D counterpart does not resolve this feedback. While the \textsc{MESA} model shows evidence that it is resolving the convective-reactive event, the shell continues to entrain fresh silicon from the upper convective boundary for longer, leading to a longer shell burning phase, albeit with lower TKE.

\begin{figure}
\centering
    \subfloat{\includegraphics[width=\columnwidth]{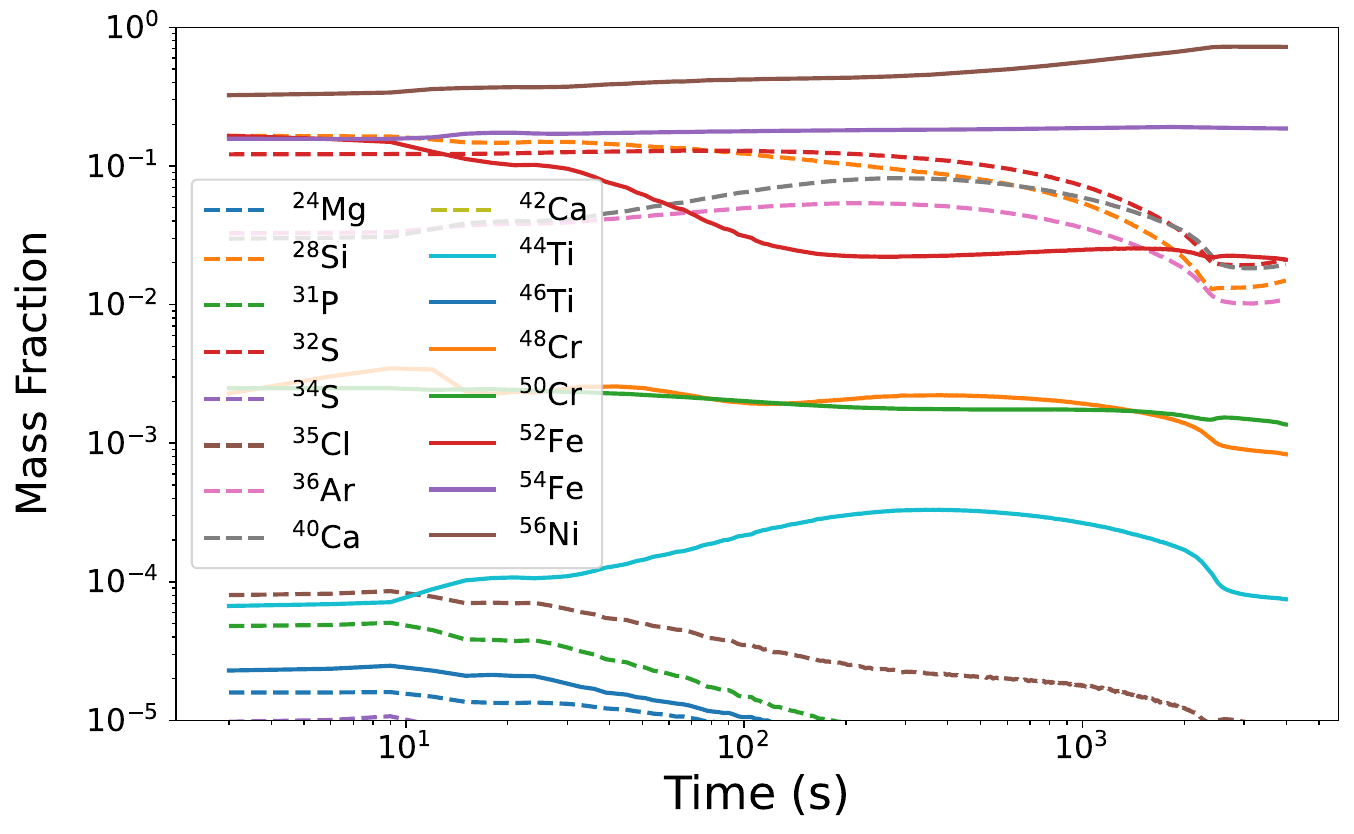}}
\hfil
    \subfloat{\includegraphics[width=\columnwidth]{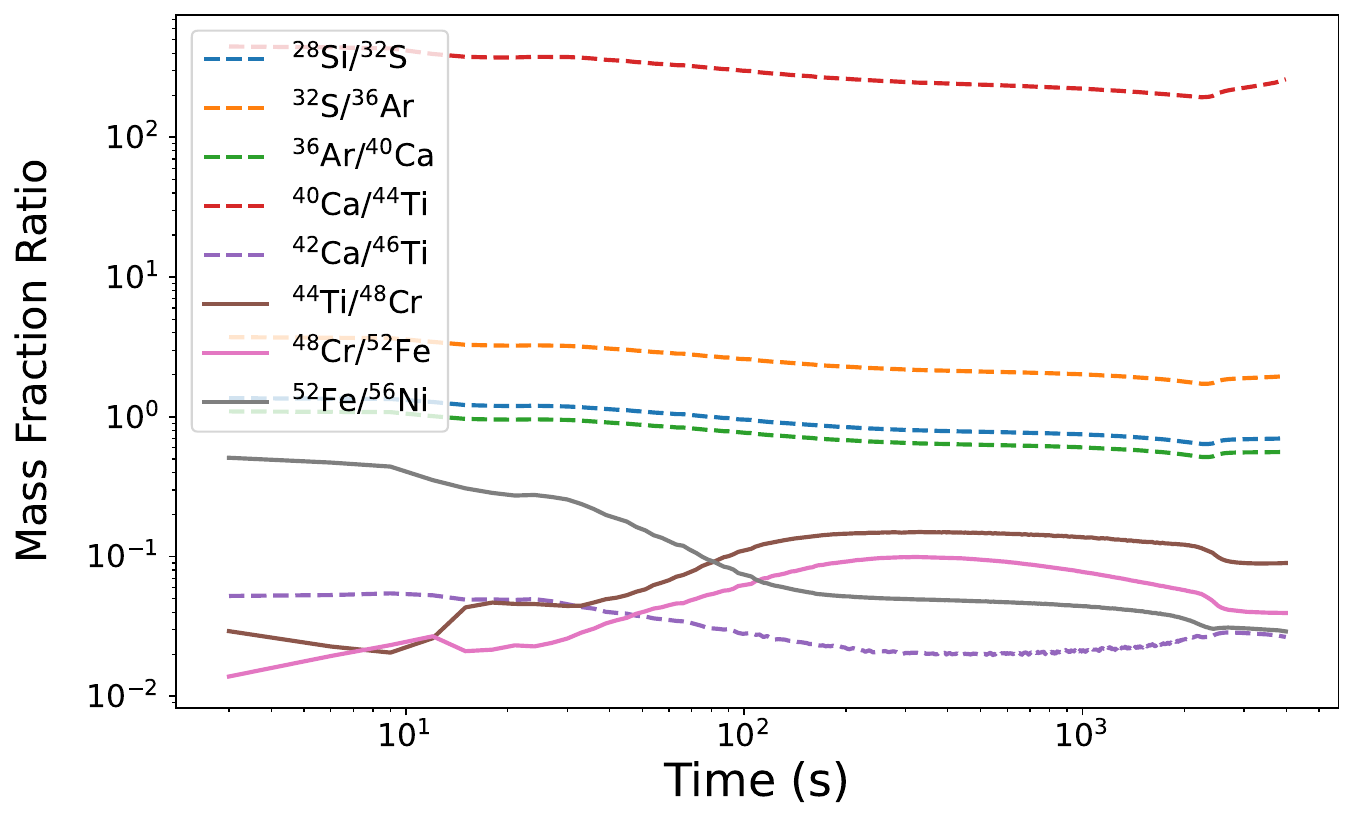}}

\caption{\normalsize Isotopes mass fractions and ratios evolution at the first energy generation peak at $1.95\times10^8$\,cm of the \texttt{45med} model.}
    \label{fig:iso_time}
\end{figure}

\section{Silicon burning and energy generation}
\label{subsec:enuc}

\begin{figure*}
\centering
    \begin{subfigure}{0.475\linewidth}
        \includegraphics[width=\linewidth]{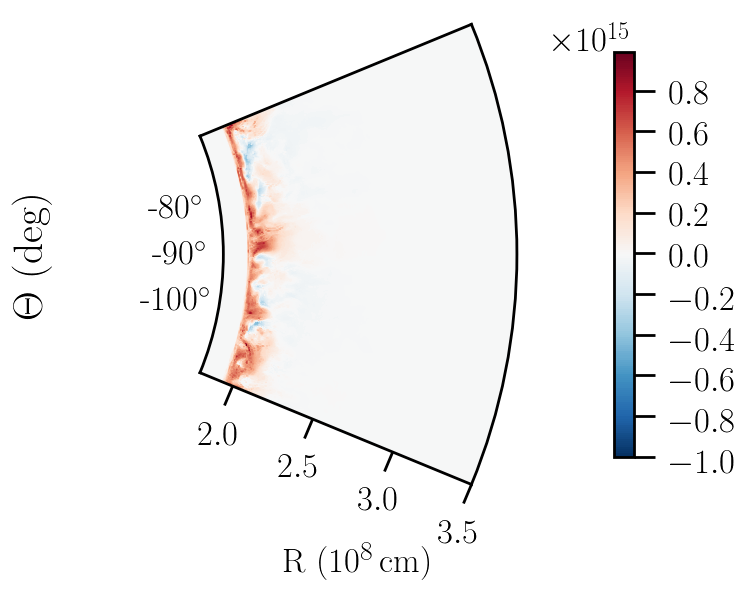}
    \end{subfigure}
\hfil
    \begin{subfigure}{0.475\linewidth}
        \includegraphics[width=\linewidth]{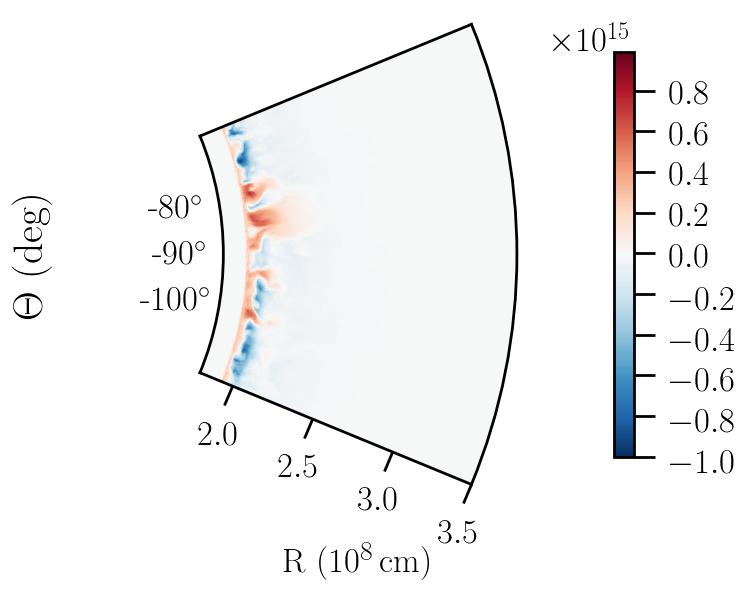}
    \end{subfigure}

    \caption{\normalsize 2D slices through model \texttt{45high} of energy generation rate at 100\,s and 1000\,s} 
    \label{fig:Enuc_2D}
\end{figure*}

\begin{figure*}
\centering
    \begin{subfigure}{0.475\linewidth}
        \includegraphics[width=\linewidth]{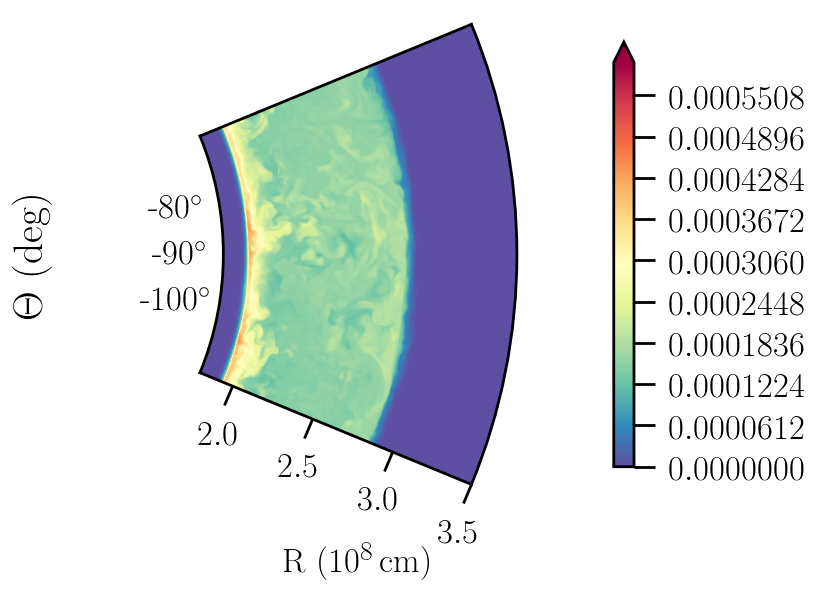}
    \end{subfigure}
\hfil
    \begin{subfigure}{0.475\linewidth}
        \includegraphics[width=\linewidth]{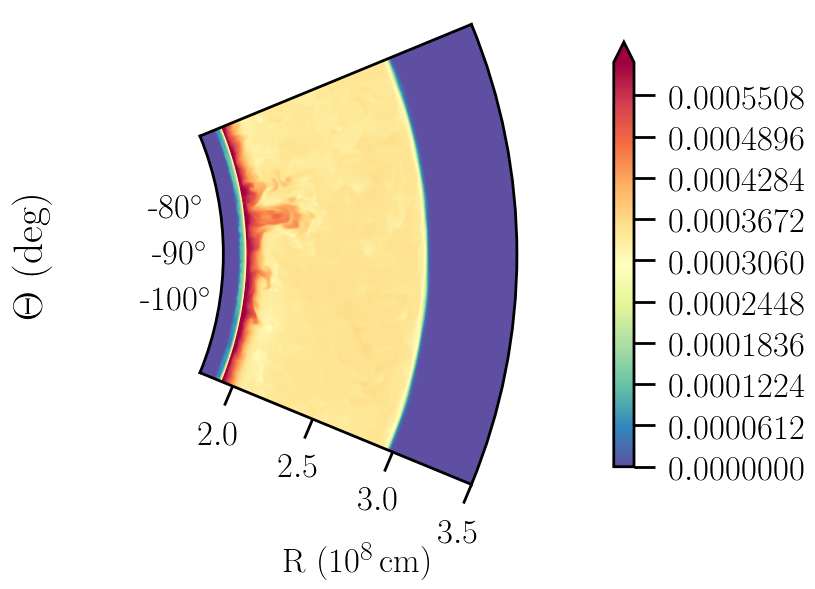}
    \end{subfigure}

    \caption{\normalsize 2D slices through model \texttt{45high} of the proton mass fraction at 100\,s and 1000\,s} 
    \label{fig:Proton_2D}
\end{figure*}

We now turn to the nuclear evolution of the silicon-burning shell and its connection to the energy-generation profile. Silicon burning differs from lighter advanced burning stages because the composition does not evolve through a simple sequence of isolated reactions. At temperatures near the base of the convective shell, a combination of capture reactions, photodisintegration, and weak interactions drive subsets of nuclei toward quasi-statistical equilibrium (QSE). From one-zone calculation studies, \citep{Hix1996,Hix1998,Hix2007}, we know that silicon burning should proceed through reaction flows between QSE groups rather than as a single linear $\alpha$-capture chain. In our reduced network, silicon-group material is represented primarily by nuclei around $^{28}\mathrm{Si}$ - $^{40}\mathrm{Ca}$, while iron-group material is represented by species such as those around $^{48}\mathrm{Cr}$ - $^{56}\mathrm{Ni}$. The intermediate nuclei, $^{40,\, 42}\mathrm{Ca}$ and $^{44,\, 46}\mathrm{Ti}$, provide a useful diagnostic of the reaction flow linking the two groups.

The upper plot of Figure~\ref{fig:iso_time} shows the time evolution of selected mass fractions measured near the base of the convection zone for the \texttt{45med} model. Although this diagnostic samples only one radial region, we find qualitatively similar behaviour throughout the active burning shell. After mapping the composition into our 3D simulation, it takes several tens of seconds for the QSE groups to form. We see this more clearly in the lower plot of Figure~\ref{fig:iso_time}, where we show the abundance ratios of nuclei along the $\alpha-$chain from $^{28}\mathrm{Si}$ to $^{56}\mathrm{Ni}$. Each isotope is shown as a ratio to its neighbouring $\alpha$ element. While these do not cover the full range of isotopes and their reactions, a flat profile of these ratios roughly indicates the balance of $\alpha$-captures and their inverse photodisintegrations. We can see in this ratio plot that between $\approx$ 100\,s and 2000\,s, these ratios are all roughly flat, indicating that silicon burning is progressing stably via the QSE groups.  We note that for our reduced-network, silicon-burning also involves proton captures, $\alpha$-induced proton emission, and reverse photodisintegration reactions, which are not visualised in these ratio plots.

As expected the silicon-group abundances, in particular $^{28}\mathrm{Si}$ and $^{32}\mathrm{S}$, decline as burning proceeds, while iron-group material builds up. The dominant iron-group product in the 3D calculation is $^{56}\mathrm{Ni}$. This is expected because the material in our hydrodynamic network remains close to $Y_e=0.5$. For matter with a small neutron excess, QSE and NSE-like arguments favour symmetric iron-group nuclei, especially $^{56}\mathrm{Ni}$. More neutron-rich material would instead shift the final composition toward species such as $^{54}\mathrm{Fe}$ and $^{58}\mathrm{Ni}$ \citep{Hix1996}. The difference between the dominant $^{56}\mathrm{Ni}$ production in our 3D model and the larger $^{54}\mathrm{Fe}$ abundance in the MESA model is primarily a consequence of the different nuclear networks and electron-fraction evolution, rather than an effect of more realistic mixing. 

The plot of mass fractions in Figure~\ref{fig:iso_time} shows that aside from the iron group elements, between 50\,s and 700\,s, we find a slight increase in the abundances of $^{36}\mathrm{Ar}$, $^{40}\mathrm{Ca}$ and $^{44}\mathrm{Ti}$. We performed a one-zone\footnote{We use the \textsc{pynucastro} package for this analysis \citep{pynucastro2, pynucastro_zenodo}.} analysis around this radius to further investigate the cause of this build-up. While it is difficult to fully disentangle the many dependent reactions during silicon burning, we notice that $^{44}\mathrm{Ti}(\alpha, \gamma)^{48}\mathrm{Cr}$ is an incredibly slow reaction. At the temperatures and densities of our burning shell, this reaction is orders of magnitude slower than $^{40}\mathrm{Ca}(\alpha, \gamma)^{44}\mathrm{Ti}$ and its inverse reaction $^{44}\mathrm{Ti}(\gamma, \alpha)^{40}\mathrm{Ca}$. This, along with reactions around it, leads to a bottleneck and a build-up of $^{44}\mathrm{Ti}$. We note that this has a feedback, as the reactions $^{40}\mathrm{Ca}(\alpha, \gamma)^{44}\mathrm{Ti}$ and $^{36}\mathrm{Ar}(\alpha, \gamma)^{40}\mathrm{Ca}$ both have remarkably similar reaction rates to their inverse reactions. This means that small perturbations in temperature can rapidly change if the forward or reverse reactions are dominant. The build-up of $^{44}\mathrm{Ti}$, thus, also leads to a build-up of $^{40}\mathrm{Ca}$ and $^{36}\mathrm{Ar}$. This qualitative behaviour holds true at larger radii, albeit at lower temperatures and hence rates.

We note that many of the rates controlling the Ca--Ti reaction flow remain experimentally uncertain or partly dependent on statistical-model estimates. Many important reactions in this range are known to be uncertain and have been largely investigated in the context of CCSNe nucleosynthesis \citep{Nassar2006, Subedi2020, Cousins2026}, as they are important in determining the final abundances of $^{44}\mathrm{Ti}$ and $^{56}\mathrm{Ni}$. Many of the reaction rates which are uncertain in this regime are proton and $\alpha$ modulated reactions \citep{Hoffman2010, Mohr2015}, which are the most important reactions in our silicon burning model. 

Our interpretation of the build-up of $^{36}\mathrm{Ar}$, $^{40}\mathrm{Ca}$, and $^{44}\mathrm{Ti}$ should therefore be understood as a qualitative description of the reaction flow in our adopted reduced network, rather than as a robust prediction of the detailed Ca--Ti abundance pattern.

Between $\approx$1000\,s and 1500\,s, we see the silicon rich material starts to deplete at a faster rate. This can be tied back to the increase in $Ri_b$ around this same time we find in Figure~\ref{fig:RIB}, which in turn, substantially decreases the entrainment rate of fresh silicon rich fuel into the convective shell. This leads to a runaway effect, quickly depleting the remaining fuel until the energy generation rate can no longer drive turbulent convection. After $\approx$2000\,s, the energy generation rate can no longer sustain turbulent convection as we have seen in Figure~\ref{fig:2Dmaps_rad}. Note, however, that since the temperatures and pressures are still very high, we continue to see evolution of the nuclear species in Figure~\ref{fig:iso_time} although there is a clear break in the trends during silicon burning.  

\begin{figure}
\centering
    \subfloat{\includegraphics[width=\columnwidth]{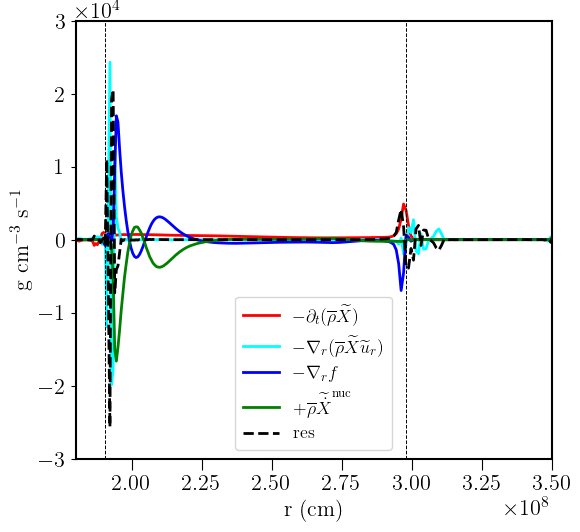}}
\hfil
    \subfloat{\includegraphics[width=\columnwidth]{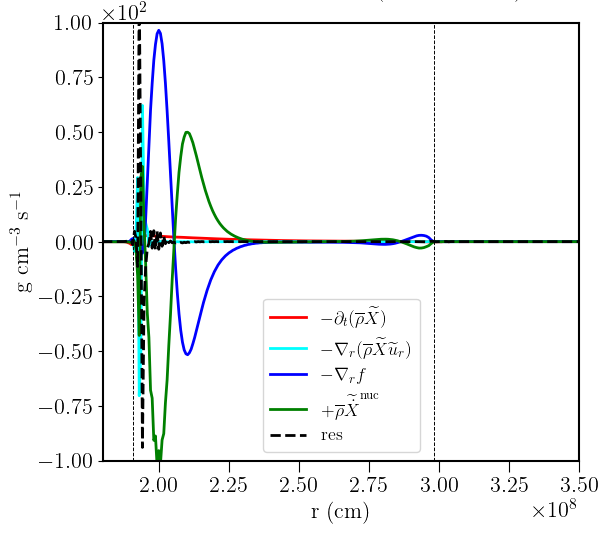}}
\hfil
    \subfloat{\includegraphics[width=\columnwidth]{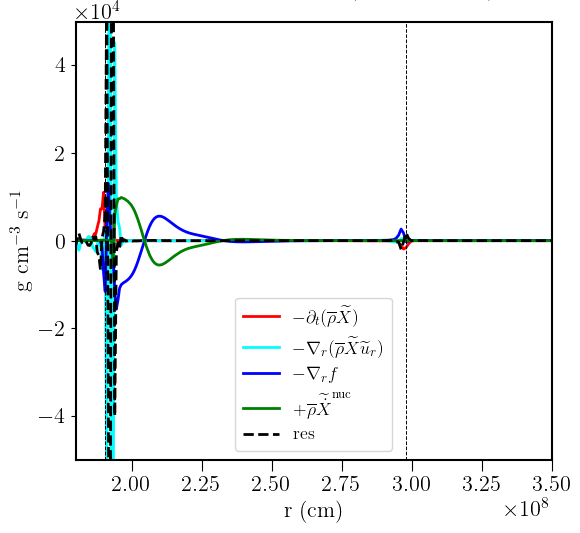}}
\caption{\normalsize Spatial and temporal averaged terms from the composition equation of $\mathrm{^{28}Si}$, $\mathrm{^{44}Ti}$, and $\mathrm{^{54}Fe}$. This plot is of the model \texttt{45high} and has been averaged over 10 convective turnovers. }
    \label{fig:RANS_comp}
\end{figure}

An important feature of the simulation is the development of a double-peaked angular averaged nuclear energy-generation profile, already shown in Figure~\ref{fig:enuc_profiles}. The profile consists of a strong positive peak near the base of the convection zone, followed by a narrow region where the net energy generation is small or slightly negative, and then a second, weaker positive peak at larger radius. A qualitatively similar structure is present in the corresponding MESA profile at early times. However, the secondary peak occurs at a somewhat larger radius in the 3D simulation, reflecting the different thermodynamic and compositional structure produced by multidimensional mixing.

It is interesting, then, that the 3D simulation maintains a stable double-peaked profile, despite having similarly sensitive reaction rates. The two-dimensional slices of the energy generation rate in Figure~\ref{fig:Enuc_2D} (a slice through model \texttt{45high}) show that the double-peaked mean profile is the angular average of a more complicated local structure. The energy generation is not organized into smooth, spherically symmetric burning layers. Instead, positive energy generation is interspersed with endoergic regions, and the pattern evolves in a flame-like manner as turbulent convection mixes the material. Similar alternating behaviour in the energy generation rate can be seen in the silicon shell of the 2D simulations of late-stage burning by \citet{Arnett2011}. 

In our 3D model, we find that the multidimensional evolution and structure of the energy generation rate is determined partly by the local proton abundance. We show this in Figure~\ref{fig:Proton_2D}, where the two 2D slices are of the proton mass fraction at the same time and location as in Figure~\ref{fig:Enuc_2D}. We find that, in particular, the regions where we see a high fraction of protons correlates well with an increase in positive energy generation rates. During silicon burning, protons are produced by photodisintegration and by reactions such as $(\alpha,p)$, and they are consumed by proton captures that can help transfer reaction flow between neighbouring nuclei and between $\alpha-$chains. 

A one-zone analysis \citep{pynucastro2, pynucastro_zenodo} along the radial trajectories within these convective zones indicates that during the steady burning phase, the dominant reactions are not $(\alpha, \gamma)$, which is what we find in the reduced single $\alpha$-chain network (approx21) for our \textsc{MESA} model, but instead $(\alpha, p)$ reactions and their inverse rates through the odd-Z elements in our nuclear network. We note that in approx21 (and similar reduced networks), $(\alpha, p)$ reactions are done implicitly.  At the base of the convection zone, this is dominated by $^{31}\mathrm{P}(p, \alpha)^{28}\mathrm{Si}$. Similar to our discussion above regarding Ti and Ca rates, the inverse reaction $^{28}\mathrm{Si}(\alpha, p)^{31}\mathrm{P}$ is usually the second fastest reaction, so small perturbations in temperature can easily cause the burning processes in the local region to be locally endothermic and endoergic. This naturally produces the alternating positive and negative energy-generation patches seen in the 2D slices. The dominance of $(\alpha, p)$ over $(\alpha, \gamma)$ reactions also explains the high abundance of protons in the convection zone compared to the \textsc{MESA} evolution (see Figures~\ref{fig:MESA_iso_full} and ~\ref{fig:PROMPI_iso_full}).

To diagnose the coupling between burning and turbulent transport more quantitatively, we apply the same Reynolds/Favre averaging framework used in Section~\ref{subsec:velocity_field} to the composition equation. For each isotope $i$, the averaged transport equation is

\begin{equation}
\partial_t(\overline{\rho}\widetilde{X}_i)
+ \nabla_r(\overline{\rho}\widetilde{X}_i\widetilde{u}_{r})
  = -\nabla_r f_i
+ \overline{\rho}\widetilde{\dot{X}}^{\rm nuc}_i ,
  \label{eq_chem}
  \end{equation}
  where $X_i$ is the mass fraction,
  \begin{equation}
  f_i = \overline{\rho X_i'' u_r''}
  \end{equation}
is the turbulent composition flux, and $\dot{X}^{\rm nuc}_i$ is the rate of production or destruction of isotope $i$ by nuclear reactions. The right-hand side of Eq.~\ref{eq_chem} separates the effects of turbulent redistribution and local nuclear processing. We focus on three representative isotopes: $^{28}\mathrm{Si}$ for the silicon-group, $^{44}\mathrm{Ti}$ for the intermediate species linking the silicon and iron groups, and $^{54}\mathrm{Fe}$ as an iron-group species.

\begin{figure}
\centering
    \subfloat{\includegraphics[width=\columnwidth]{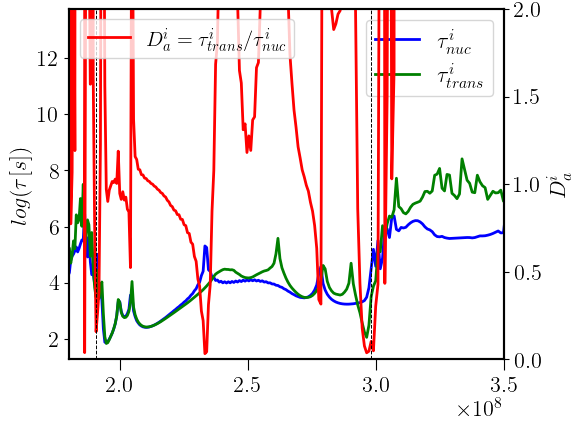}}
\hfil
    \subfloat{\includegraphics[width=\columnwidth]{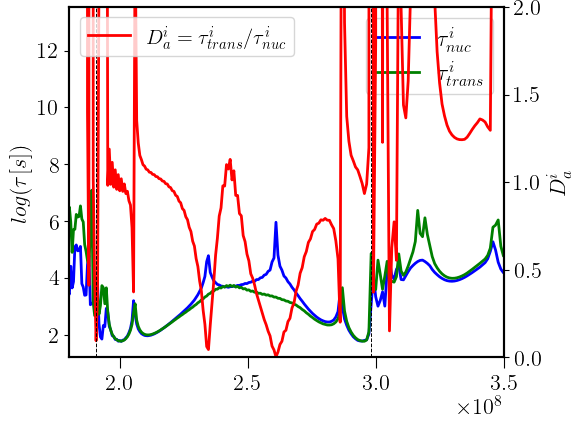}}
\hfil
    \subfloat{\includegraphics[width=\columnwidth]{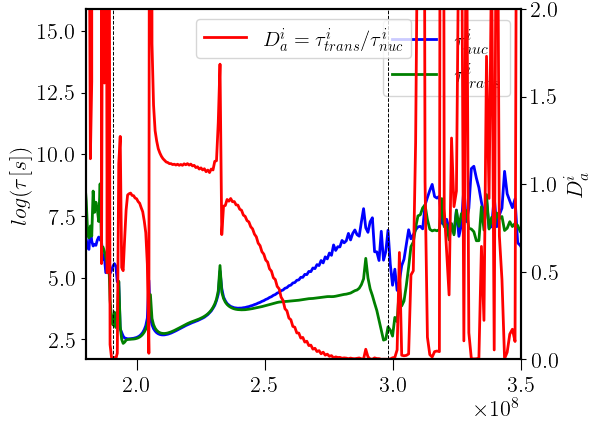}}
\caption{\normalsize Nuclear and transport timescales alongside Damkohler number plots of $\mathrm{^{28}Si}$, $\mathrm{^{44}Ti}$, and $\mathrm{^{54}Fe}$. This plot is of the model \texttt{45high} and has been averaged over 10 convective turnovers.}
    \label{fig:DA_comp}
\end{figure}

Similar to Figure~\ref{fig:TKE}, both Figure~\ref{fig:RANS_comp} and Figure~\ref{fig:DA_comp} are of the \texttt{45high} model, and we have temporally averaged over three convective turnovers between a physical time of 850\,s and 1000\,s.
In Figure~\ref{fig:RANS_comp}, for all three species the time derivative, $\partial_t(\overline{\rho}\widetilde{X}_i)$, is largely negligible. However, we see a small increase at the upper convective boundary, where some fresh $^{28}\mathrm{Si}$ is being entrained, and a similar increase at the lower convective boundary for $^{54}\mathrm{Fe}$ due to the production of iron-group elements here. 

The mean composition flux term, $\nabla_r(\overline{\rho}\widetilde{X}_i\widetilde{u}_{r})$, is also largely negligible throughout the domain for all three species, however, we see a spike for all of them around the lower convective boundary. This indicates a change in the composition profile here, due to their destruction in this region for $^{28}\mathrm{Si}$ and $^{44}\mathrm{Ti}$, and the production of $^{54}\mathrm{Fe}$. In previous convective shell studies, \citep{Georgy2024, mocak_18}, both the time derivative and mean composition flux terms have been shown to be non-negligible at the upper convective boundary when fresh material of that species is entrained. We find little indication of that in these simulations, however, due to the stiff convective boundaries limiting the mass entrainment. The residual term in each of these plots also mirror our mean composition flux term, indicating that our numerical dissipation is highest where we encounter rapid variations. Previous studies have shown that the residual can be decreased by increasing our grid resolution \citep{Arnett2019}.

Figure~\ref{fig:RANS_comp} also shows the nuclear source terms
$\overline{\rho}\widetilde{\dot{X}}^{\rm nuc}_i$ for $^{28}\mathrm{Si}$,
$^{44}\mathrm{Ti}$, and $^{54}\mathrm{Fe}$. The $^{28}\mathrm{Si}$ source term is approximately anti-correlated with the nuclear energy-generation profile. In the innermost burning peak, $^{28}\mathrm{Si}$ is destroyed and the net energy generation is positive. In the intermediate region, where the averaged energy generation becomes small or slightly negative, the nuclear source term for $^{28}\mathrm{Si}$ becomes weakly positive. This indicates partial regeneration of silicon-group material, most likely through reverse reactions and QSE readjustment of material that has been mixed away from the hottest part of the shell. At larger radii, the term becomes negative again, corresponding to the secondary positive energy-generation peak and destruction of silicon-group material.

The behaviour of $^{44}\mathrm{Ti}$ supports this interpretation. As an intermediate species, $^{44}\mathrm{Ti}$ is depleted in the lower part of the shell where the reaction flow proceeds efficiently toward the iron group. At larger radii, where the temperature is lower and the reaction flow is less complete, $^{44}\mathrm{Ti}$ can accumulate. The transition from $^{44}\mathrm{Ti}$ destruction to production therefore marks the radial region where the flow between the silicon and iron QSE groups becomes less efficient. This is consistent with our discussion above, noting that there is a bottleneck in the reaction rates at $^{44}\mathrm{Ti}$.

The $^{54}\mathrm{Fe}$ source term shows a similar pattern to $^{44}\mathrm{Ti}$. It is produced in the deeper burning region, where silicon-group material is processed into iron-group ash. At larger radii, however, the $^{54}\mathrm{Fe}$ source term becomes negative. Since the global electron fraction remains close to $Y_e=0.5$, $^{54}\mathrm{Fe}$ is not the preferred final iron-group product in our network. Its destruction at larger radii likely reflects continued QSE rearrangement toward more symmetric iron-group material, especially $^{56}\mathrm{Ni}$, together with reverse reactions in regions where the material has not reached complete equilibrium. We note that $^{56}\mathrm{Ni}$ does not follow the same pattern as we have seen with $^{44}\mathrm{Ti}$ and $^{54}\mathrm{Fe}$\footnote{See Appendix for additional composition RA-ILES plots for proton, $\alpha$, $^{12}\mathrm{C}$ and $^{56}\mathrm{Ni}$}. Instead, it is just a single peaked structure which decreases gradually to zero at around $2.4\times10^8\mathrm{cm}$. 

For all species, we also find that the divergence of the turbulent flux term, $\nabla_r f_i$, directly opposes the nuclear source term due to the replenishment of fuel. Unlike with previous stable stellar burning phases, where the nuclear timescale is always expected to be reasonably long, silicon burning can occur through $\alpha$ and proton-captures which can be much faster. To visualise the comparison between the nuclear and timescales in this silicon shell, we compute the Damk\"ohler number for each of the same three isotopes as above, and plot them in Figure~\ref{fig:DA_comp},
\begin{equation}
Da = \frac{\tau_{\rm mix}}{\tau_{\rm nuc}},
\qquad
\tau_{\rm nuc,i} =
\left| \frac{\widetilde{X}}
{\widetilde{\dot{X}}^{\rm nuc}} \right| ,
\label{eq:damkohler}
\end{equation}
where $\tau_{\rm mix}$ is estimated as the ratio between the convective length scale, $\ell_c = \Delta r_{conv}$, and the rms velocity. Regions with $Da\gtrsim1$ are convective-reactive, where nuclear processing occurs on a timescale comparable to, or shorter than, turbulent mixing. Compared to previous studies of deep convective shells, the plots of Da for all species during silicon burning is very messy and complicated. However, we note that aside from the inner several $100\,$km of the convective shell, both the nuclear and transport timescales in the outer regions are often much longer than the lifetime of the silicon burning shell. Focusing on just the inner regions, we still find that for all three nuclear species, the nuclear timescale is often shorter or equal to that of the transport timescale throughout the important burning regions, corresponding to the double peaked energy generation profiles. This leads to the composition gradients shown in Figure~\ref{fig:PROMPI_iso_full}. We also find similar convective-reactive events during shell merger and interaction events \citep{mocak_18, Ritter2018, Rizzuti2024}.

We therefore interpret the double-peaked energy-generation profile as a consequence of convective-reactive silicon burning in a QSE-regulated network. The first peak corresponds to the efficient destruction of silicon-group fuel near the hot base of the convection zone. The intermediate negative region is produced by local QSE readjustment, reverse reactions, and partial regeneration of silicon-group material. The second positive peak occurs where mixed fuel is processed again, but under different thermodynamic conditions and with a different local mixture of light particles and intermediate nuclei. Because the flow is turbulent, these processes do not occur in smooth spherical layers. Instead, the averaged profile emerges from the angular superposition of hot, fuel-rich, ash-rich, proton-rich, exoergic, and endoergic patches.

This behaviour reinforces the conclusion that the shell is in a convective-reactive regime. Silicon-rich material exists above the burning layer, but entrainment across the convective boundaries is too slow to replenish the fuel at the rate at which it is consumed. Within the convection zone itself, the Damk"ohler numbers and composition source terms show that nuclear processing and turbulent mixing occur on comparable timescales. The abundance structure and energy-generation profile therefore cannot be understood from the temperature profile alone; they depend on the coupled evolution of turbulent transport, QSE reaction flow, and local proton and $\alpha$ abundances.

\section{Conclusions}
\label{sec:conclusion}
For the first time, we have simulated a complete convective silicon burning shell in 3D until shell exhaustion for a non-rotating $14\,\mathrm{M_\odot}$ progenitor from \citet{Whitehead2026}. We used the three-dimensional hydrodynamic code \textsc{PROMPI} with a reduced 25-isotope nuclear network. Our aim has been to study how multidimensional turbulent mixing affects the evolution of active silicon-burning, and how this differs from the behaviour of the corresponding one-dimensional MESA model.

The broad evolution of the silicon-burning shell in 3D is similar to the main high-TKE phase of the 1D model. After an initial transient following the mapping from MESA to \textsc{PROMPI}, the shell enters a quasi-steady convective-burning phase with typical rms velocities of order $10^7,\mathrm{cm\,s^{-1}}$. However, the 3D silicon-burning shell dies earlier than its 1D counterpart. We find that this is primarily due to differences in convective boundary mixing. In the MESA model, the imposed diffusive CBM prescription eventually connects the active silicon-burning shell to a previously convective layer above it, supplying fresh silicon-rich material and extending the lifetime of the shell. Despite these differences, it is clear from Figure~\ref{fig:2Dmaps_rad} and Figure~\ref{fig:2Dmaps_mass} that qualitatively, our 3D simulation still matches the 1D stellar evolution model very well. 

This weaker entrainment in 3D is linked to the stiffness of the convective boundaries. In the 3D simulation, the CBM naturally depends on the physical conditions of the convective boundaries and the layer beyond them. The upper boundary remains very stiff, limiting the entrainment across this boundary. The bulk Richardson numbers of both the inner and outer boundaries are large throughout most of the simulation, indicating that the turbulent motions do not have enough kinetic energy to efficiently overcome the stabilising buoyancy barriers. As the shell begins to run out of fuel, the nuclear energy generation drops, the convective velocity decreases, and the bulk Richardson number rises further. This produces a feedback in which declining burning weakens the convection, weaker convection further suppresses entrainment, and the shell ultimately exhausts its available fuel. In this model, therefore, the effective boundary mixing in 3D is weaker than that implied by the 1D overshoot prescription. This is an important indication that for 1D stellar evolution models, applying a single CBM value across all convective boundaries is not sufficient. Future improvements should instead determine CBM based on the physical conditions at each convective boundary individually (e.g. using $Ri_b$).

We have also tested the sensitivity of the flow to resolution and angular opening angle. The bulk properties of the convective shell are similar across our models, suggesting that the main conclusions are not strongly dependent on the resolutions considered here. The larger $90^\circ$ domains show a stronger horizontal velocity component, which may be related to the thinness of the shell and the stiffness of its boundaries. A similar effect has also been noted in the silicon shell of the rotating simulations of \citet{Griffiths2026}. In their simulation, this shell also has very high bulk Richardson numbers. 

The turbulent kinetic energy budget supports the picture of a quasi-steady, weakly entraining convective shell during the main burning phase. Buoyancy work drives turbulence throughout most of the convective region, while the stable layers near the boundaries decelerate radial motions. Turbulent and acoustic fluxes redistribute kinetic energy within the shell, and the residual term acts as a sink associated with numerical dissipation at the grid scale. The overall balance of these terms shows that the flow remains statistically steady until the fuel supply begins to fail. The nuclear evolution shows that silicon burning in this shell is strongly convective-reactive. With our reduced 25-isotope network, we find indications that the burning proceeds through QSE-like groups as expected. The abundance evolution is not described by a simple $\alpha$-capture chain, with important contributions from reverse reactions, proton captures, and $(\alpha,p)$ reactions. The dominant iron-group product in the 3D model is $^{56}\mathrm{Ni}$, which we attribute mainly to the reduced network and the fact that the material remains close to $Y_e=0.5$. We reiterate that many proton and $\alpha$ modulated reactions which we identify as the most important set of reactions have large experimental uncertainties \citep{Nassar2006, Hoffman2010, Mohr2015, Subedi2020, Cousins2026}.

A distinctive feature of the simulation is the double-peaked angular-averaged energy-generation profile. The first peak corresponds to efficient destruction of silicon-group material near the hot base of the convective shell, quickly forming Fe and Ni isotopes. The intermediate region, where net energy generation becomes small or negative, appears to be associated with reverse reactions and QSE readjustment, including partial regeneration of the silicon-group material. A second positive peak occurs farther out, where the mixed fuel is processed again under different thermodynamic and compositional conditions. Here, due to the lower temperatures, only partial silicon burning is active, up to material around $^{44}\mathrm{Ti}$. In two-dimensional slices this structure is not smooth, but consists of interspersed exoergic and endoergic regions that evolve as the turbulent flow advects fuel, ashes, protons, and $\alpha$ particles through the shell.

In the model studied here, silicon-rich material remains available above the active shell, but the convective boundaries are too stiff for this material to be entrained quickly enough to sustain burning. At the same time, the Damk\"ohler numbers and the composition source terms show that nuclear processing and turbulent mixing occur on comparable timescales within the shell itself. The abundance structure and energy-generation profile therefore cannot be understood from the temperature profile alone, but depend on the full multidimensional convective-reactive evolution.

There are several important limitations to this first study. The reduced 25-isotope network is sufficient for following the broad energetics and reaction-flow behaviour, but it cannot provide final detailed nucleosynthetic yields. The treatment of weak interactions and the resulting electron-fraction evolution are especially important for the final iron-group composition. Future work will therefore need to repeat this analysis with larger nuclear networks and more complete weak-interaction physics. A caveat to this is that since many of the reaction rates during silicon burning are extremely sensitive to slight differences/fluctuations in thermodynamic variables. Mapping a 1D model to a 3D simulation with substantially different nuclear networks can lead to a quick loss of hydrostatic equilibrium and even significant changes in the location of the main burning fronts. An improved study of a more complete nuclear network in 3D will therefore require an improved network to also be used in the 1D stellar evolution calculations as well. Nevertheless, this work demonstrates that multidimensional turbulent mixing can qualitatively affect the lifetime, boundary evolution, and local nuclear burning structure of convective silicon shells.

\section*{Acknowledgements}

VV acknowledges support from the Royal Astronomical Society Research Fellowship. 
RH acknowledges support from STFC, the World Premier International Research Centre Initiative (WPI Initiative), MEXT, Japan, the IReNA AccelNet Network of Networks (NSF, Grant No. OISE-1927130), CeNAM (grant DE-SC0026204) and {the Wolfson Foundation that part-funded the greenHPC facility at Keele}.
This work used the DiRAC Memory Intensive service (Cosma8) at Durham University, managed by the Institute for Computational Cosmology on behalf of the STFC DiRAC HPC Facility (www.dirac.ac.uk). The DiRAC service at Durham was funded by BEIS, UKRI and STFC capital funding, Durham University and STFC operations grants. DiRAC is part of the UKRI Digital Research Infrastructure. 
FR is a fellow of the Alexander von Humboldt Foundation, and acknowledges support by the Klaus Tschira Foundation and by the INAF Mini grant 2024, ‘GALoMS -- Galactic Archaeology for Low Mass Stars’ (1.05.24.07.02).

\section*{Data Availability}
The data underlying this article will be shared on reasonable request to the  authors, subject to considerations of intellectual property law.



\bibliographystyle{mnras}
\bibliography{paper} 



\appendix

\section{Composition RA-ILES}
Here we present a plot of the spatial and temporally averaged RA-ILES composition equation as described in Section~\ref{subsec:enuc}. These plots are effectively the same as what was presented and discussed in Figure~\ref{fig:RANS_comp}, but given for protons, $\alpha$, $\mathrm{^{12}C}$ and $\mathrm{^{56}Ni}$. These additional plots may help give a more complete picture of what has already been discussed, especially since many of the most important reactions that we've discussed have been proton and $\alpha$-modulated reactions. $\mathrm{^{12}C}$ gives an indication of the behaviour of some of the low mass isotopes which undergo both photodisintegration into $\alpha$ and $\alpha$ captures, and the $\mathrm{^{56}Ni}$ composition plot shows the production of $\mathrm{^{56}Ni}$ at the base of the silicon burning region, as it is the main product of this burning process in our 3D simulation. 

\begin{figure*}
\centering
    \begin{subfigure}{0.4\linewidth}
        \includegraphics[width=\linewidth]{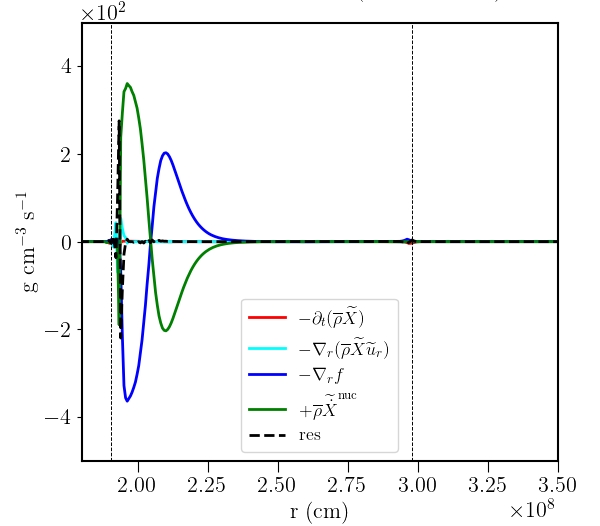}
    \end{subfigure}
\hfil
    \begin{subfigure}{0.4\linewidth}
        \includegraphics[width=\linewidth]{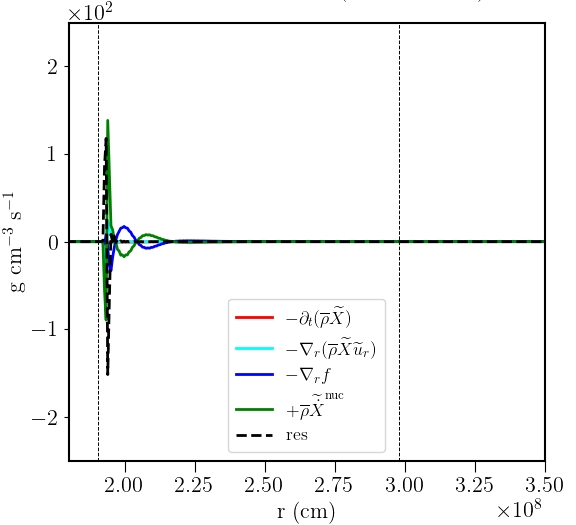}
    \end{subfigure}

    \begin{subfigure}{0.4\linewidth}
        \includegraphics[width=\linewidth]{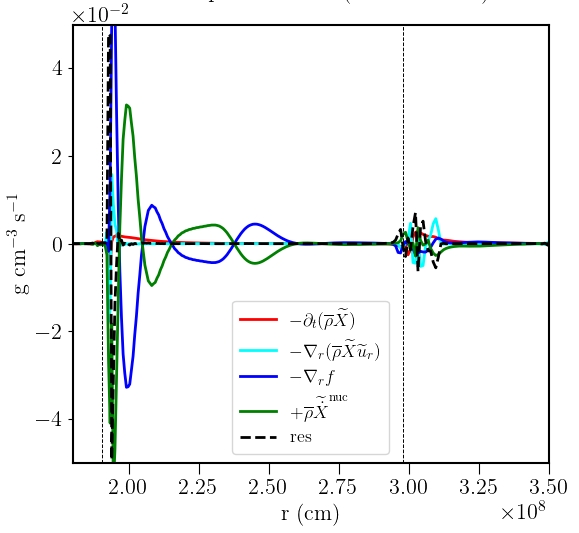}
    \end{subfigure}
\hfil
    \begin{subfigure}{0.4\linewidth}
        \includegraphics[width=\linewidth]{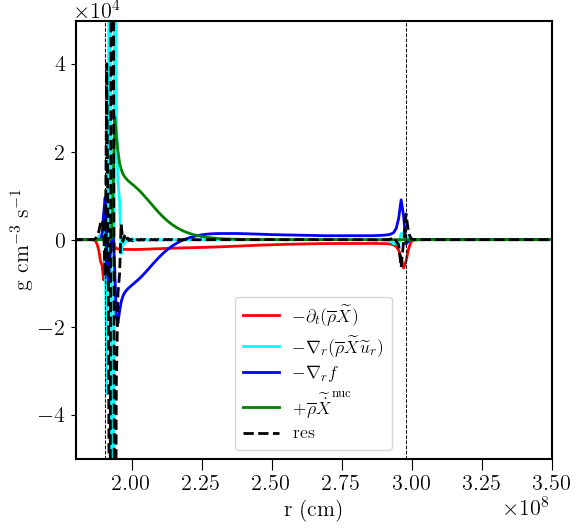}
    \end{subfigure}
\caption{\normalsize Spatial and temporal averaged terms from the composition equation of protons (top left), $\mathrm{^{4}He}$ (top right), $\mathrm{^{12}C}$ (bottom left), and $\mathrm{^{56}Ni}$ (bottom right). These plots are of the model \texttt{45high} and has been averaged over 10 convective turnovers. }
    \label{fig:RANS_comp_app}
\end{figure*}


\bsp	
\label{lastpage}
\end{document}